\documentclass[aps,prl,twocolumn,amsmath,amssymb,nofootinbib,superscriptaddress,floatfix]{revtex4-1}

\usepackage{amsmath,amsfonts,amssymb}
\usepackage{amssymb}
\usepackage{amsthm}
\usepackage[dvipdf]{color}
\usepackage{graphicx}
\usepackage{dcolumn}
\usepackage{bm} 
\usepackage{ulem}
\usepackage{cancel}

\begin{document}
\title{Mechanical Origin of High-Temperature Thermal Stability in Platinum Oxides}
	
\author{Fangyuan Ma}
\thanks{These authors contribute equally to this work.}
%\affiliation{Key Lab of Advanced Optoelectronic Quantum Architecture and Measurement (MOE), School of Physics, Beijing Institute of Technology, Beijing, 100081, China}
\affiliation{Beijing Key Laboratory of Quantum Matter State Control and Ultra-Precision Measurement Technology, School of Physics, Beijing Institute of Technology, Beijing, 100081, China}

\author{Mengzhao Sun}
\thanks{These authors contribute equally to this work.}
\affiliation{School of Physical Science and Technology \& Shanghai Key Laboratory of High-resolution Electron Microscopy, ShanghaiTech University, Shanghai, 201210, 
China}
\affiliation{Center for Transformative Science, ShanghaiTech University, Shanghai, 201210, China}

\author{Xuejian Gong}
\thanks{These authors contribute equally to this work.}
%\affiliation{Key Lab of Advanced Optoelectronic Quantum Architecture and Measurement (MOE), School of Physics, Beijing Institute of Technology, Beijing, 100081, China}
\affiliation{Beijing Key Laboratory of Quantum Matter State Control and Ultra-Precision Measurement Technology, School of Physics, Beijing Institute of Technology, Beijing, 100081, China}

\author{Jun Cai}
%\email{author4@XXX.edu.cn}
\affiliation{School of Physical Science and Technology \& Shanghai Key Laboratory of High-resolution Electron Microscopy, ShanghaiTech University, Shanghai, 201210, 
China}
\affiliation{Center for Transformative Science, ShanghaiTech University, Shanghai, 201210, China}

\author{Zhujun Wang}
\email{wangzhj3@shanghaitech.edu.cn}
\affiliation{School of Physical Science and Technology \& Shanghai Key Laboratory of High-resolution Electron Microscopy, ShanghaiTech University, Shanghai, 201210, 
China}
\affiliation{Center for Transformative Science, ShanghaiTech University, Shanghai, 201210, China}

\author{Di Zhou}
\email{dizhou@bit.edu.cn}
%\affiliation{Key Lab of Advanced Optoelectronic Quantum Architecture and Measurement (MOE), School of Physics, Beijing Institute of Technology, Beijing, 100081, China}
\affiliation{Beijing Key Laboratory of Quantum Matter State Control and Ultra-Precision Measurement Technology, School of Physics, Beijing Institute of Technology, Beijing, 100081, China}

\begin{abstract}
Platinum oxides are vital catalysts, but their limited thermal stability hinders applications. Recent studies have uncovered a structural transition in two-dimensional platinum oxides that significantly enhances their thermal resilience by several hundred Kelvin. Herein, we demonstrate that this enhanced stability stems from the mechanical robustness of the elastic network at the atomic scale. Prior to the transition, an over-constrained lattice generates localized states of self-stress through an incommensurate Moir\'{e} pattern with the platinum substrate, reducing thermal endurance. After the transition, the oxide shifts to a mechanically flexible structure with balanced degrees of freedom and constraints. The isostatic network, together with the platinum substrate, forms a commensurate Moir\'{e} superlattice that relaxes elastic energy and enhances stability. These findings highlight the fundamental role of network connectivity in governing thermal stability, and provide a design principle for catalysts in extreme environments.
\end{abstract}

\maketitle

\textit{Introduction---}Platinum oxides are key catalysts in chemical processes, including fuel cell technology~\cite{park2000direct}, automotive catalysis~\cite{van2017observing}, and organic synthesis~\cite{wilson2014synthetic}. However, their broader application is limited by poor thermal stability at a few hundred Kelvin~\cite{steininger1982adsorption}. For instance, when exposed to nitrogen dioxide~\cite{devarajan2008stm}, platinum surfaces form an out-of-plane dice-lattice oxide structure~\cite{van2017observing, weaver2005oxidation}, which decomposes near $700\,\text{K}$ and thereby limits their utility~\cite{vanSpronsen2017NC}.

Recent advances show that on the platinum (111) surface, this out-of-plane dice lattice of platinum oxide undergoes a structural transition above the conventional melting point, forming a distinct six-pointed star structure~\cite{wang2024NM}. This newly identified phase exhibits exceptional thermal stability even at $1200$ K, challenging the prevailing view that platinum oxides remain unstable at only a few hundred Kelvin~\cite{chaston1965reactions}. Although this enhanced thermal stability expands the catalytic performance of platinum oxides, its physical mechanism remains unclear. Furthermore, whether this structural design principle can be generalized to other chemical compounds remains an open question.

In this work, we show that the thermal stability of platinum oxides is governed by an intrinsic index of network topology~\cite{maxwell1864xlv, Lubensky2015RPP} of the elastic structure modeled by atomic nodes connected by chemical bonds. This topological index, namely the number of self-stress states, is defined by atomic degrees of freedom versus chemical constraints, and it exhibits remarkable distinction before and after the structural transition.

Prior to the transition, the oxide dice lattice is over-constrained, with more constraints than atomic degrees of freedom. Such lattices inherently host mechanical self-stress states~\cite{tan2025classifying,zhang2018fracturing}, originating from bond-length fluctuations in practical fabrications. At the atomic scale, these variations stem from horizontal misalignment between oxygen and platinum atoms in the underlying (111) surface, which induces interlayer bonding. Because the oxide dice lattice is incommensurate with the platinum (111) surface~\cite{he2021moire}, the resulting Moir\'e pattern imposes quasi-periodic modulation of bond lengths, localizing self-stress states at maximal variation.

%After the transition, a six-pointed star structure of platinum oxide emerges through removal of platinum atoms from the parent dice lattice. This transformation reduces constraints relative to degrees of freedom, evolving the system to \textit{mechanically isostatic}~\cite{Wyart2005,Zhang2009N,Fruchart2020N,Lei2021PRL}, where constraints and degrees of freedom are perfectly balanced. This star lattice lies at the verge of mechanical instability~\cite{kane2014NP,Zhou2019PRX,Bertoldi2017}, reducing self-stress states to a sub-extensive number when supported on the platinum (111) substrate. Moreover, this flexible structure conforms to the substrate, yielding a commensurate Moir\'{e} superlattice~\cite{mcgilly2020visualization}. This spatial periodicity enables interlayer bond-length variations to adopt a regular pattern, thereby reducing the localization profile of self-stress states. 

After the transition, a six-pointed star structure of platinum oxide emerges through removal of platinum atoms from the parent dice lattice. This transformation reduces constraints relative to degrees of freedom. As a result, the system evolves to a \textit{mechanically isostatic} state~\cite{Wyart2005,Zaccone5,Zhang2009N,Fruchart2020N,Lei2021PRL, Zaccone2}, where constraints and degrees of freedom are perfectly balanced. The star lattice lies at the verge of mechanical instability~\cite{kane2014NP,Zhou2019PRX,Bertoldi2017, Zaccone3}, and when supported on the platinum (111) substrate, self-stress states are reduced to a sub-extensive number. Moreover, this flexible structure conforms to the substrate, yielding a commensurate Moir\'{e} superlattice~\cite{mcgilly2020visualization}. This spatial periodicity regularizes interlayer bond-length variations, suppressing the spatial concentration of self-stress states.

We demonstrate the thermal stability of platinum oxides through stochastic Newtonian dynamics based on the Langevin equation. In the over-constrained dice lattice, an extensive number of highly localized stresses confine elastic energy near incommensurate Moir\'{e} positions, where interlayer bond-length fluctuations are maximal. As temperature increases, these stresses intensify rapidly, triggering thermal instability. In contrast, the commensurate Moir\'{e} pattern formed by the flexible star lattice exhibits a sub-extensive number of extended tensions, enabling a more uniform distribution of elastic energy under thermal excitation and allowing the star-structured platinum oxide to remain thermally stable.

This framework reveals that network connectivity-rather than geometry or chemical composition-governs the thermal robustness of compounds such as platinum oxides. Our findings open new avenues for designing materials that are either thermally stable or intentionally unstable, spanning length scales from atomic to macroscopic.

\begin{figure}
	\centering
	\includegraphics[scale=0.41]{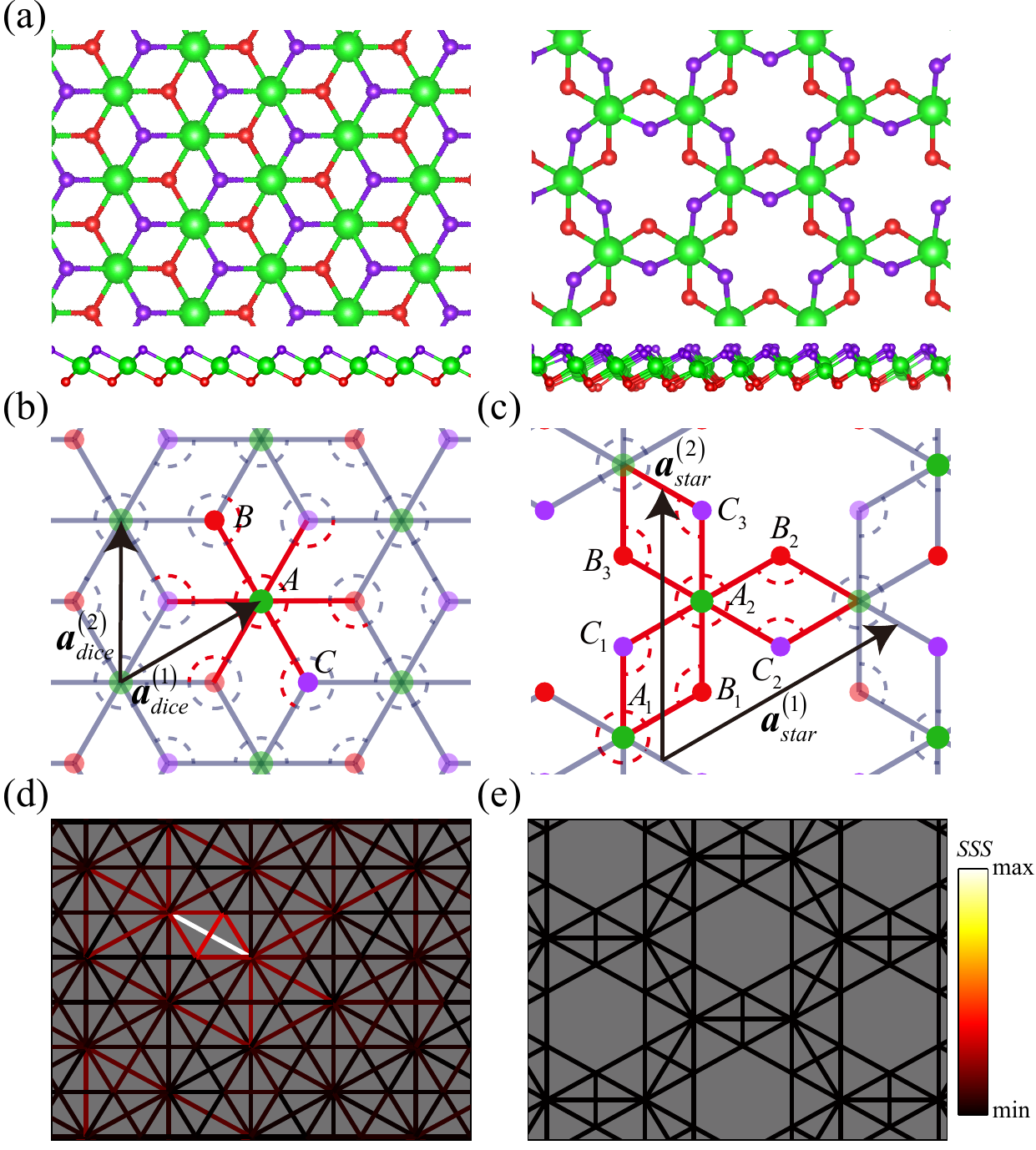}
	\caption{Elastic network modeling of two-dimensional platinum oxide sheets based on \textit{in situ} experiments~\cite{wang2024NM}. 
    %\textcolor[rgb]{0,0,1}{(a) Atomic structures of platinum oxide dice-lattice (left) and the six-pointed star-lattice (right). In the side view, oxygen atoms at the vacuum interface (purple) and near the platinum layer (red) indicated.} 
    %(b) Unit cell of the dice lattice, containing a platinum atom (green dot) at $\bm{A} = \ell_0(0,0,2.3)$ and lower/upper oxygen atoms (red and purple dots) at $\bm{B} = \ell_0(\cos \tfrac{2\pi}{3}, \sin \tfrac{2\pi}{3}, 1.71)$ and $\bm{C} = \ell_0(\cos \tfrac{5\pi}{3}, \sin \tfrac{5\pi}{3}, 2.85)$, respectively. Here $\ell_0 = 1.726\,\text{\AA}$ denotes the horizontal projection of the nearest-neighbor edge length, indicated by solid red lines common to both lattices. Bending constraints are shown as red arcs. Primitive vectors are indicated by black arrows: $\bm{a}_{\rm dice}^{(1)} = \ell_0(\sqrt{3}\cos \tfrac{\pi}{6}, \sqrt{3}\sin \tfrac{\pi}{6}, 0)$ and $\bm{a}_{\rm dice}^{(2)} = \ell_0(0,\sqrt{3},0)$. 
  (a) Atomic structures of dice-lattice (left) and six-pointed star-lattice (right). In side view, oxygen atoms near the vacuum interface (purple) and near the platinum layer (red) are indicated. (b) Unit cell of the dice lattice, containing a platinum atom (green) at $\bm{A}=\ell_0(0,0,2.3)$ and oxygen atoms (red, purple) at $\bm{B}=\ell_0(\cos \tfrac{2\pi}{3}, \sin \tfrac{2\pi}{3},1.71)$ and $\bm{C}=\ell_0(\cos \tfrac{5\pi}{3}, \sin \tfrac{5\pi}{3},2.85)$. Here $\ell_0=1.726\,\text{\AA}$ is the horizontal projection of the nearest-neighbor edge length (solid red lines). Bending constraints are shown as red arcs. Primitive vectors are indicated by black arrows: $\bm{a}_{\rm dice}^{(1)}=\ell_0(\sqrt{3}\cos \tfrac{\pi}{6}, \sqrt{3}\sin \tfrac{\pi}{6},0)$ and $\bm{a}_{\rm dice}^{(2)}=\ell_0(0,\sqrt{3},0)$.
(c) Unit cell of the star lattice, hosting two platinum atoms at 
$\bm{A}_1 = \ell_0(\sqrt{3}\cos \tfrac{4\pi}{3}, \sqrt{3}\sin \tfrac{4\pi}{3}, 2.3)$ and 
$\bm{A}_2 =\ell_0(0,0,2.3)$. 
Each platinum atom is coordinated by six oxygen atoms. 
Lower and upper oxygen atoms (red and purple dots) are located at 
$\bm{B}_i = \ell_0(\cos \tfrac{(4i-7)\pi}{6}, \sin \tfrac{(4i-7)\pi}{6}, 1.71)$ and 
$\bm{C}_i = \ell_0(\cos \tfrac{(4i-9)\pi}{6}, \sin \tfrac{(4i-9)\pi}{6}, 2.85)$ for $i=1,2,3$. 
Primitive vectors are 
$\bm{a}_{\rm star}^{(1)} = \ell_0(3\cos \frac{\pi}{6}, 3\sin \frac{\pi}{6}, 0)$ and 
$\bm{a}_{\rm star}^{(2)} = \ell_0(0,3,0)$. 
(d) Spatial profile of a self-stress state in the dice lattice, with color indicating stress intensity. 
Self-stress in bonds and bending constraints is represented by edges and diagonal lines. 
(e) The star lattice, being mechanically isostatic, exhibits no self-stress states.}
\label{fig1}
\end{figure}

% derived from the rank-nullity theorem of the compatibility matrix~\cite{}, 

\textit{Network connectivity and topological index of self-stress states in platinum oxides---}We first analyze the out-of-plane dice and star lattices as two-dimensional sheets embedded in three-dimensional space, while deferring discussion of the bilayer system, which is formed by placing oxide structures atop the platinum (111) substrate, to the next section. Each atom is treated as a mass point, and every nearest-neighbor platinum-oxygen bond is modeled as a Hookean spring. Experimental evidence and density functional theory~\cite{Vashishta1990PRB, Chen2019FOP, PhysRevMaterials.7.075601} indicate that each hinge (e.g., the Pt-O-Pt and O-Pt-O angles) possesses finite bending energy, which we encode as a bending constraint. The connectivity of an elastic network is commonly characterized by the Maxwell-Calladine counting principle~\cite{maxwell1864xlv, paulose2015topological, Ma2023PRL, tan2025classifying, Zaccone4, zaccone1}, through the integer
\begin{eqnarray}
N_{\rm sss} = N_{\rm c} - nd,
\end{eqnarray}
where $N_{\rm c}$ is the number of constraints and $nd$ the degrees of freedom in the unit cell, with $n$ atoms and dimensionality $d=3$. $N_{\rm sss}$ gives the number of self-stress states per unit cell. Such states, widely studied in elastic networks, describe internal force patterns without external load~\cite{PhysRevLett.116.135501, baardink2018localizing, paulose2015topological, Zhou2018PRL, Lubensky2015RPP, Mao2025PRL}. Importantly, $N_{\rm sss}$ is a network topological invariant: it remains unchanged under continuous geometric deformations and only alters when chemical bonds break and structural transition occurs.

In the dice lattice, the unit cell contains one platinum and two oxygen atoms, giving $nd = 9$ degrees of freedom. The two oxygen atoms occupy inequivalent out-of-plane positions, one higher and the other lower relative to the platinum plane. As shown in Figs.~\ref{fig1}(a,b), six bonds together with 12 bending constraints impose $N_{\rm c} = 18$ constraints per unit cell, yielding $N_{\rm sss} = 9$. Since $N_{\rm sss} > 0$, the dice lattice is over-constrained and mechanically rigid.

In contrast, the star lattice achieves a perfect balance between degrees of freedom and constraints, placing it at the verge of mechanical instability. Its unit cell contains two platinum and six oxygen atoms, three oxygen atoms higher and three lower relative to the platinum plane, as shown in Figs.~\ref{fig1}(a,c), giving $nd = 24$ degrees of freedom. The vertices, stabilized by bending rigidities that suppress angular distortions, form three rigid tetrahedra per unit cell~\cite{tang2024PRL}. Within each tetrahedron, these bending constraints can be equivalently represented as diagonal bonds~\cite{Lubensky2019PRL}. Each unit cell comprises 12 nearest-neighbor bonds, 6 diagonal constraints within the tetrahedra, and 6 bending rigidities between neighboring tetrahedra, yielding $N_{\rm c} = 24$. These constraints exactly balance the $24$ degrees of freedom, giving $N_{\rm sss} = 0$ and placing the star lattice at the mechanical \textit{isostatic} point.

%Unlike the over-constrained dice lattice, which hosts an extensive number of self-stress states (Fig.~\ref{fig1}(d)), the star lattice is mechanically flexible and devoid of such states (Fig.~\ref{fig1}(e)). The integer index $N_{\rm sss}$, defined for the two-dimensional oxide sheets, qualitatively distinguishes the mechanical and thermal stability of the dice and star lattices. In the following sections, when these oxide sheets are placed atop the platinum substrate, the same index continues to signal their sharply different mechanical responses, independent of microscopic parameters. In what follows, we adopt the parameters in Figs.~\ref{fig1} and \ref{fig2} to demonstrate the qualitatively distinct mechanical responses of the bilayer structures.

%The integer index $N_{\rm sss}$ captures the essential distinction in mechanical and thermal stability between platinum oxides supported on the platinum substrate, independent of specific parameters. 

\begin{figure}
	\centering
	\includegraphics[scale=0.42]{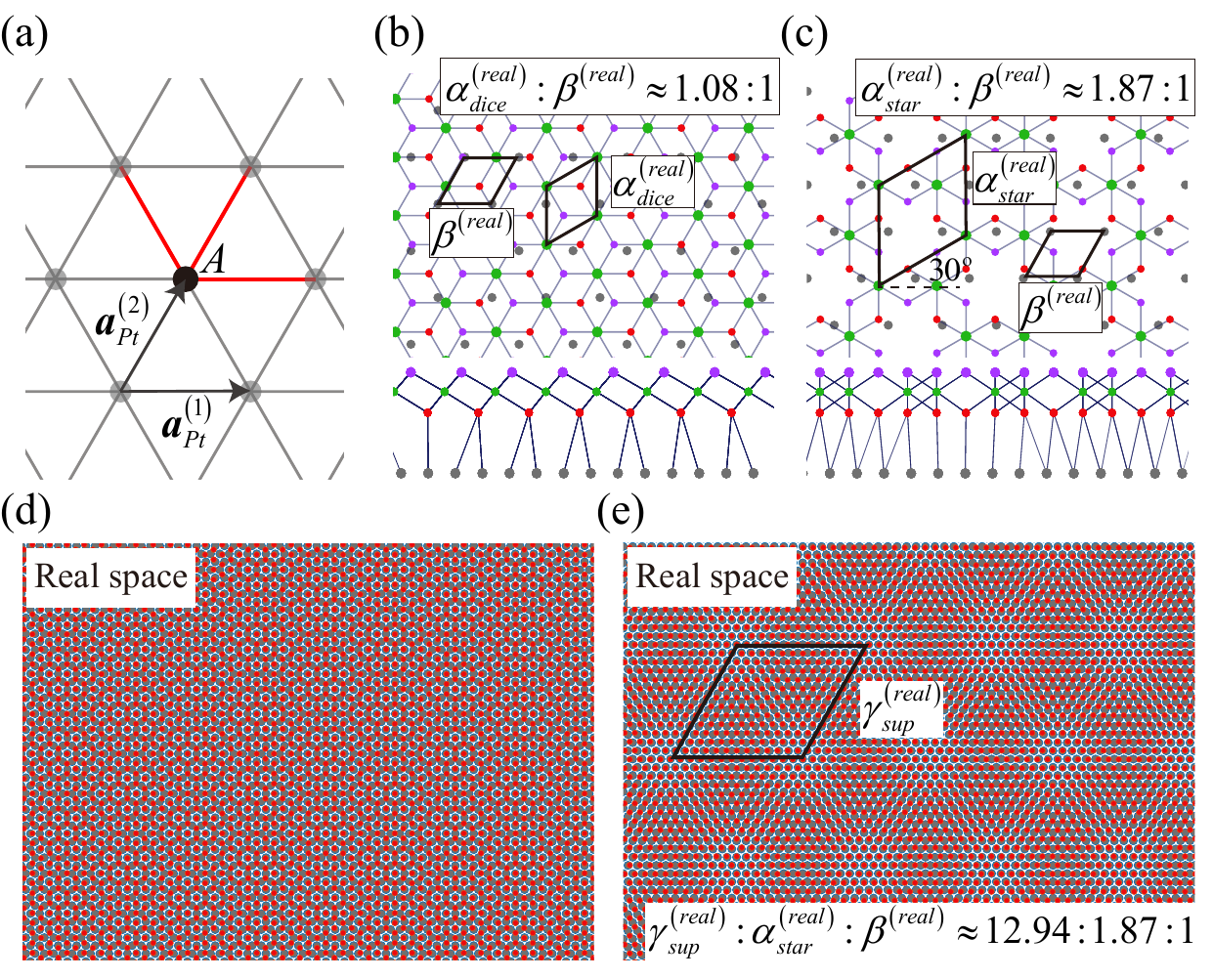}
    \caption{Initial bilayer structures prior to mechanical relaxation. (a) Geometry of the substrate: The primitive vectors of the triangular surface lattice are $\bm{a}_{\mathrm{Pt}}^{(1)} = \ell_0(1.606, 0, 0)$ and $\bm{a}_{\mathrm{Pt}}^{(2)} = \ell_0\bigl(1.606\cos\frac{\pi}{3},1.606\sin\frac{\pi}{3},0\bigr)$. The top platinum (111) layer (height $= 0$) evolves dynamically, while the subsurface layer (height $= -1.311\ell_0$) is fixed to model the rigid bulk of the platinum substrate. 
(b,c) Top views of the dice (b) and star (c) bilayers on the substrate. The rhombuses $\alpha_{\mathrm{dice}}^{(\mathrm{real})}$ and $\alpha_{\mathrm{star}}^{(\mathrm{real})}$ mark oxide unit cells; $\beta^{(\mathrm{real})}$ denotes the substrate unit cell. Side views (bottom) show the bonding of the lowest oxygen atoms (red dots) to the substrate platinum atoms.
(d,e) Arrangement of the lowest oxygen atoms (oxide layer) and the top platinum atoms (substrate layer) forming interlayer bonds, generating (d) an incommensurate Moir\'{e} quasicrystal for the dice bilayer and (e) a commensurate Moir\'{e} superlattice for the star bilayer. The superlattice unit cell is outlined by black rhombus $\gamma_{\mathrm{star}}^{(\mathrm{real})}$ in (e).
}\label{fig2}
\end{figure}

Unlike the over-constrained dice lattice, which hosts an extensive number of self-stress states (Fig.~\ref{fig1}(d)), the star lattice is flexible and devoid of them (Fig.~\ref{fig1}(e)). This qualitative distinction of the oxide sheets is captured by the index $N_{\rm sss}$. When the sheets are later placed atop the platinum substrate, this index continues to reveal their contrasting mechanical responses, independent of microscopic parameters. In what follows, we adopt the parameters in Figs.~\ref{fig1} and \ref{fig2} to demonstrate these bilayer behaviors.

% defined for the oxide sheets, already captures this qualitative distinction. 

\textit{Zero-temperature mechanical stability of platinum oxides on a platinum substrate---}We construct dice and star bilayer structures by placing the oxide lattices on the platinum (111) surface, which forms a triangular lattice (Fig.~\ref{fig2}(a)). To model the interaction with the substrate, the top platinum (111) layer directly in contact with the oxide evolves dynamically, while the immediate subsurface layer of platinum atoms is held fixed to capture the close-packing rigidity~\cite{crc_handbook_2024, kittel2004introduction, PhysRevB.97.224305}. Oxides are anchored to the top substrate layer via oxygen-platinum bonds: each oxygen in the oxide lower layer pairs with the nearest platinum atom within a horizontal distance of $0.721\ell_0$ ($\ell_0=1.726\text{\AA}$), as illustrated in Figs.~\ref{fig2}(b,c).

Each atom $i$ is modeled as a point mass $m_i$ at position $\bm{r}_i$. Chemical bonds between atoms $i$ and $j$ are represented as Hookean springs with stiffness $k_{ij}$ and rest length $\ell_{ij}$. For intralayer bonds, $\ell_{ij}$ equals the platinum-oxygen bond length in the oxide lattice, whereas for interlayer bonds it corresponds to the vertical separation between the lowest oxygen atom and the top platinum substrate, $1.71\ell_0$. Bending hinges are defined at vertex $j$ by adjacent nodes $i$ and $k$, each with bending stiffness $\kappa_{ijk}$ and preferred angle $\Theta_{ijk}$. Intralayer hinges reproduce the geometry of un-relaxed oxide lattices, while interlayer hinges are fixed at $\Theta_{ijk}=\pi/2$. The total elastic energy~\cite{Florijn2014, Coulais2016} is $U(\{\bm{r}\}) = \sum_{\langle i,j\rangle}k_{ij}(|\bm{r}_i-\bm{r}_j|-\ell_{ij})^2 /2
+ \sum_{\langle i,j,k\rangle}\kappa_{ijk}(\theta_{ijk}-\Theta_{ijk})^2 /2$, where $\theta_{ijk}$ denotes the instantaneous opening angle at vertex $j$. At zero temperature, the equilibrium configuration of the bilayers, $\{\bm{r}^{(0)}\}$, is obtained by minimizing $U(\{\bm{r}\})$. Because the primitive vectors of the oxide lattices are incompatible with the platinum substrate, the interlayer bonds undergo stretching and the oxide lattices relax into new mechanical equilibrium configurations. We set the stretching stiffness and bending rigidity to $k_{ij}=100~\text{N/m}$ and $\kappa_{ijk}/\ell_0^2=0.302~\text{N/m}$, respectively, with values consistent with first-principles estimates~\cite{Chen2019FOP} and reproduce experimental bilayer configurations~\cite{wang2024NM}.

\begin{figure}
	\centering
	\includegraphics[scale=0.41]{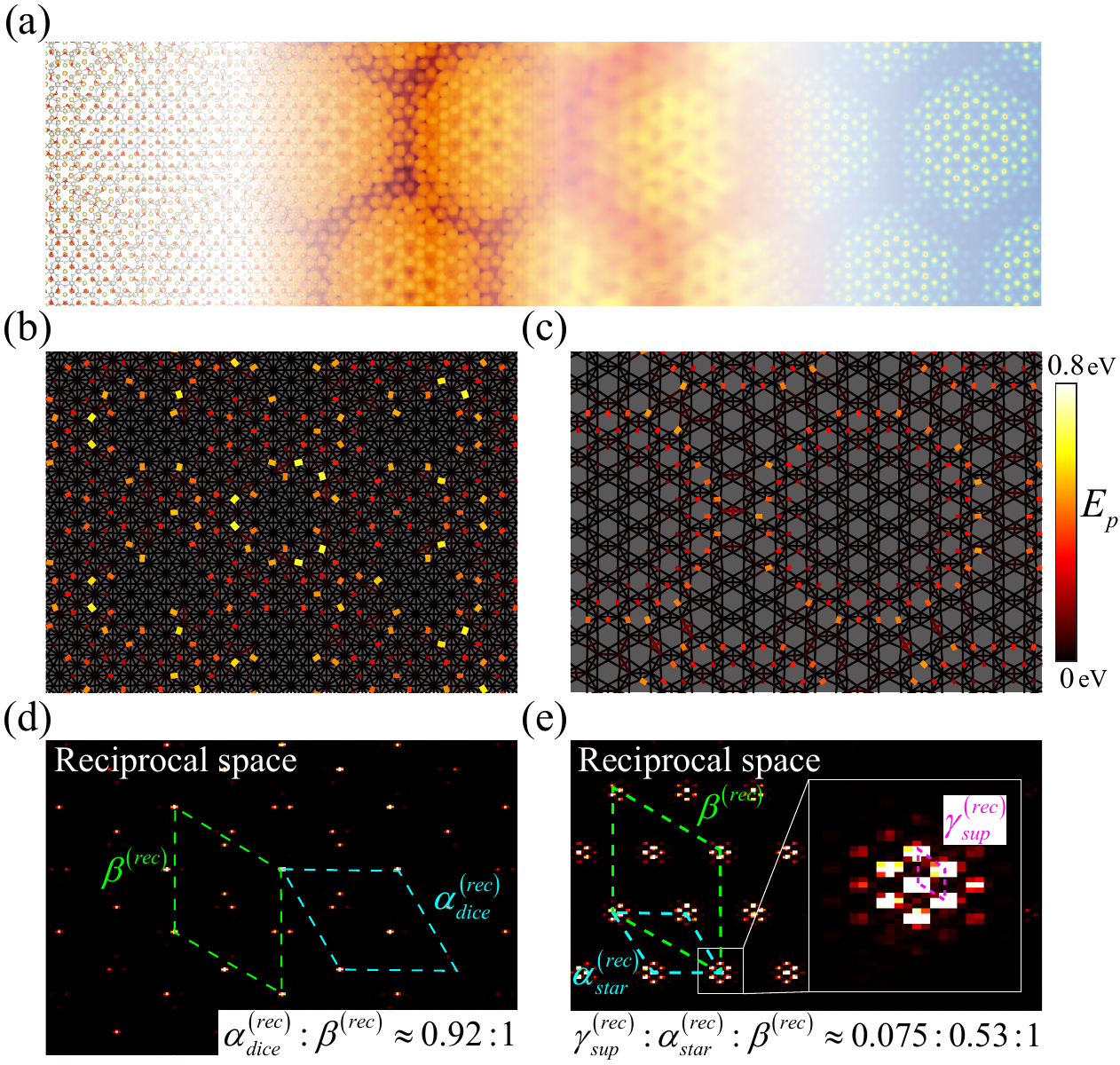}
	\caption{Bilayer structures after mechanical relaxation. (a) Platinum oxides with star lattice, shown left to right: atomic model, STM simulation, STM experiment, and electron density distribution. (b,c) Spatial profiles of elastic energy in the dice and star lattices, with colors on nearest-/next-nearest-neighbor edges indicating stretching and bending contributions. 
(d,e) Reciprocal lattices of the two Moir\'e bilayer patterns of site positions. Here, $\alpha_{\rm dice}^{(\rm rec)}$ and $\alpha_{\rm star}^{(\rm rec)}$ represent reciprocal lattice vectors of dice and star lattices, while $\beta^{(\rm rec)}$ corresponds to those of the platinum substrate. The reciprocal vector of the Moir\'e superlattice, highlighted in the enlarged square of (e), is denoted by $\gamma_{\rm star}^{(\rm rec)}$.}
\label{fig3}
\end{figure}

The over-constrained dice lattice cannot adapt to the triangular substrate, thereby generating an incommensurate~\cite{Yao2018PNAS} Moir\'e pattern. This incommensurability induces quasicrystalline interlayer bond-length fluctuations, with self-stress localized in regions of maximal length variation, as reflected by strongly fluctuating elastic energy across the Moir\'e quasicrystal in Fig.~\ref{fig3}(b). In Fig.~\ref{fig3}(d), Fourier analysis of site positions further highlights the non-crystalline character of the reciprocal lattice. At zero temperature, the dice-bilayer thus accumulates large elastic energy and becomes mechanically unstable.

By contrast, the mechanically isostatic star lattice, when anchored to the platinum substrate, supports only a sub-extensive number of self-stress states. Because of its mechanical flexibility, the star lattice conforms to substrate geometry, forming a commensurate Moir\'{e} superlattice aligned with the platinum (111) surface. This crystalline superlattice further suppresses bond-length fluctuations, producing extended self-stress with reduced amplitude, as shown in Fig.~\ref{fig3}(c).

%Numerical simulations show that self-stress forms hexagonal rings, with dark regions inside and between rings marking minimal stress and bright rings indicating maximal stress. Comparing Fig.~\ref{fig3}(c) with the third panel of Fig.~\ref{fig3}(a) (STM experiment), these dark regions correspond to the brightest and darkest areas in Fig.~\ref{fig3}(a), reflecting maximal and minimal atomic overlap. Before relaxation, they signify the shortest and longest interlayer bond lengths; after equilibrium, intermediate overlap regions undergo the strongest stretching and store the most elastic energy, while maximal and minimal overlap regions stretch least. This spatial distribution of self-stress matches Fig.~\ref{fig3}(c). Moreover, in Fig.~\ref{fig3}(e), reciprocal-space analysis reveals six-fold crystalline symmetry and a small-wavevector pattern consistent with the bilayer modulation in Figs.~\ref{fig3}(a,c). These Fourier-space features are also supported by LEED measurements~\cite{wang2024NM}, validating the network modeling of platinum oxide bilayers. At zero temperature, the spatially flattened star-bilayer accumulates only sub-extensive elastic energy, conferring notable mechanical resilience.

Numerical simulations show that self-stress organizes into hexagonal rings. Dark regions inside and between the rings mark minimal stress, while bright rings indicate maximal stress. Comparing Fig.~\ref{fig3}(c) with the STM experiment in Fig.~\ref{fig3}(a), the dark regions correspond to areas of maximal and minimal atomic overlap, whereas the bright rings align with intermediate atomic overlap regions. Before relaxation, these overlaps signify the shortest and longest interlayer bond lengths; after equilibrium, intermediate overlap regions undergo the strongest stretching and store the most elastic energy, while maximal and minimal overlap regions stretch least. Consequently, this spatial distribution of self-stress matches Fig.~\ref{fig3}(c). Moreover, in Fig.~\ref{fig3}(e), reciprocal-space analysis reveals six-fold crystalline symmetry and a small-wavevector pattern consistent with the bilayer modulation in Figs.~\ref{fig3}(a,c). These Fourier-space features are also supported by LEED measurements~\cite{wang2024NM}, validating the network modeling of platinum oxide bilayers. At zero temperature, the spatially flattened star-bilayer accumulates only sub-extensive elastic energy, conferring notable mechanical resilience.

% Moreover, in Fig.~\ref{fig3}(e), reciprocal-space analysis reveals six-fold crystalline symmetry, while the Brillouin-zone center exhibits a small-wavevector pattern consistent with the large-wavelength bilayer modulation in Figs.~\ref{fig3}(a). These Fourier-space features closely match the LEED pattern observed experimentally for the star-bilayer~\cite{wang2024NM}, thereby validating the network modeling of platinum oxide bilayers. 

\begin{figure}
\centering
\includegraphics[scale=0.38]{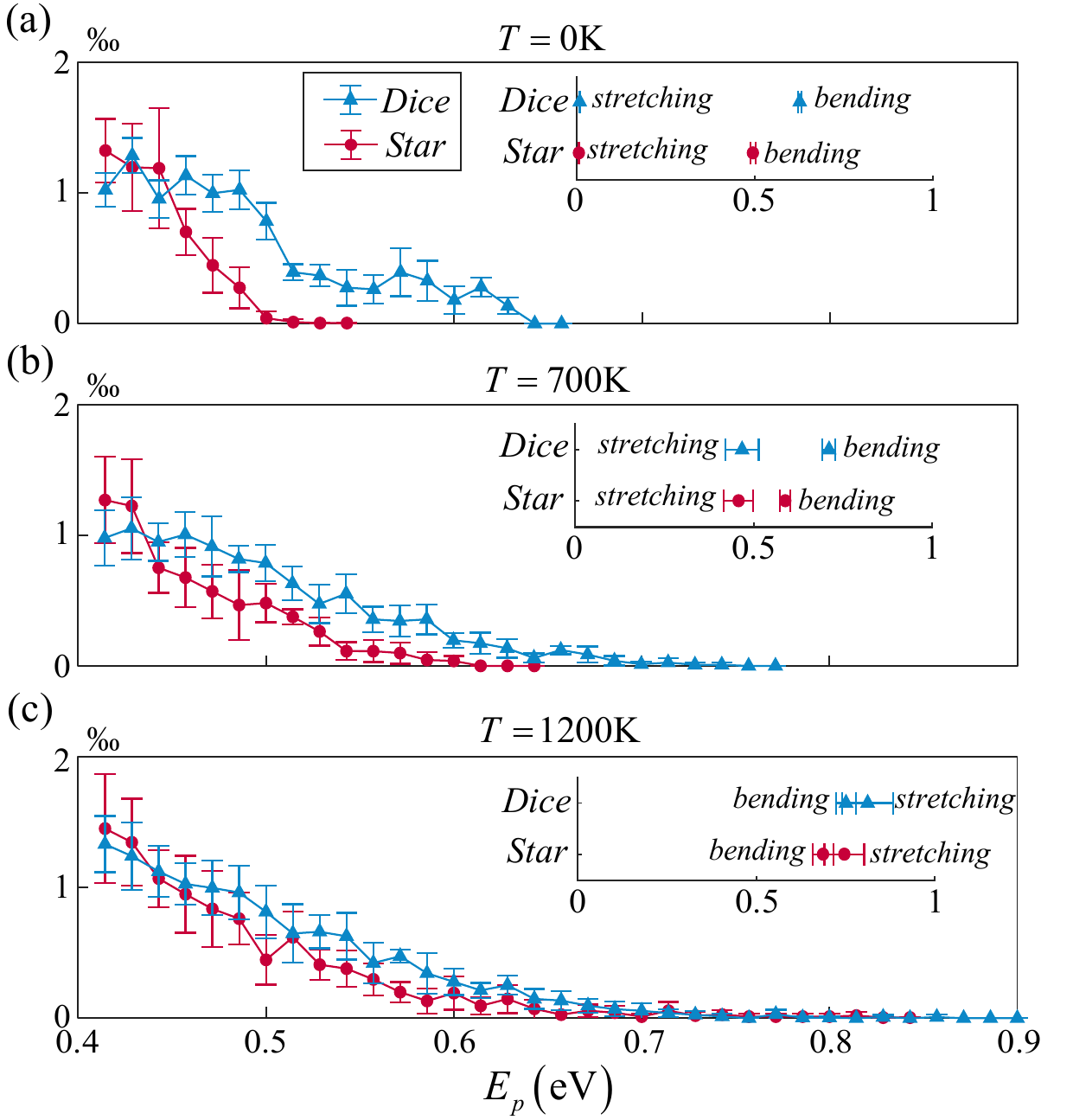}
\caption{Percentage distributions of stretching and bending energies per bond in bilayer structures 
at different temperatures: $0\,\text{K}$ in (a), $700\,\text{K}$ in (b), and $1200\,\text{K}$ in (c). 
Each bilayer is modeled as a rhombus with side length $ 104.0\,\text{\AA} = 60.25\ell_0$, comprising 
$35 \times 35$ unit cells with $2.3 \times 10^{4}$ bonds in the dice-bilayer, 
and $20 \times 20$ unit cells with $1.3 \times 10^{4}$ bonds in the star-bilayer. For statistical sampling, 10 dice-lattice and 10 star-lattice bilayers were generated by varying $\delta \bm{r}_{\rm Pt} = \xi^{(1)} \bm{a}_{\rm Pt}^{(1)} + \xi^{(2)} \bm{a}_{\rm Pt}^{(2)}$, 
which denotes the coordinate origin of the platinum substrate relative to the top platinum-oxide sheets, 
where $\xi^{(1,2)} \in [0,1]$ are uniformly distributed random variables. The horizontal axis shows elastic energies per constraint, while the vertical axis gives the normalized fraction (percentage population) of bonds/angles within each energy 
interval. Each data point corresponds to $E \pm \delta E/2$, with spacing $\delta E = 0.02\,\text{eV}$; 
the vertical axis is in units of one-thousandths. Distributions are truncated below 
$0.5\,\text{eV}$ to emphasize high-energy components (full statistics in Supplementary Information~\cite{SIPtOx}). 
Blue and red curves denote dice and star bilayers, respectively. Triangular and circular markers 
indicate mean values, with error bars showing standard deviations. Insets in (a-c) present the 
maximum stretching and bending energies per bond across the 10 random 
configurations for dice and star lattices at the corresponding temperatures.
    }\label{fig4}
\end{figure}

\textit{Finite-temperature thermal stability of bilayer structures---}We assess bilayer thermal stability under finite-temperature effects via the Langevin equation~\cite{RevModPhys.38.541, RevModPhys.49.435} 
(Newtonian dynamics with stochastic thermal driving). 
Phononic excitations arise from atomic displacements $\bm{u}_i=\bm{r}_i-\bm{r}_i^{(0)}$ from equilibrium, driven by fluctuations. 
These vibrations follow the Langevin equation,
\begin{eqnarray}\label{Langevin}
m_i\ddot{\bm{u}}_i = -\frac{\partial U(\{\bm{r}\})}{\partial \bm{u}_i} 
-\eta \dot{\bm{u}}_i + \sqrt{2\eta k_B T}\,\bm{\xi}_i(t),
\end{eqnarray}
where $m_{\rm Pt}=195.08\,\text{u}$ and $m_{\rm O}=16\,\text{u}$ are the atomic masses of platinum and oxygen, with $\text{u}$ the atomic mass unit. 
The damping coefficient is $\eta=4.07 \times 10^{-13}$ kg/s, estimated from the Sutherland viscosity law~\cite{Chapman1970, batchelor2000introduction}. 
$k_B$ denotes Boltzmann’s constant and $T$ the temperature. 
The stochastic term $\bm{\xi}_i(t)$ is white noise with correlation $\langle \bm{\xi}_i(t)\bm{\xi}_j(t')\rangle=\delta_{ij}\delta(t-t')$.

Molecular dynamics simulations, performed in non-dimensionalized units of the Langevin equation, were run until the bilayers reached thermal equilibrium~\cite{SIPtOx}. Elastic energies were then measured over 10 randomly generated bilayer structures obtained by varying relative positions between the platinum substrate and oxide lattices. Due to zero-temperature self-stress states, the statistics of finite-temperature elastic energy per bond reveal several key features in Fig.~\ref{fig4}.

At $0\,\text{K}$, the dice-bilayer shows significantly larger elastic energy per constraint than the star-bilayer. Inset statistics of the 10 largest values further illustrate this, with the dice-bilayer energy about $1.3$ times higher, consistent with the spatial distribution in Figs.~\ref{fig3}(b,c). At $700\,\text{K}$, where the dice-bilayer melts, its elastic energy per constraint exceeds that of the star-bilayer, reflecting their respective instability and stability. At $1200\,\text{K}$, the star-bilayer stores substantially more elastic energy than at lower temperatures, comparable to the dice-bilayer's melting energy at $700\,\text{K}$. This rise suggests that the star-bilayer may reach its melting point near $1200\,\text{K}$. Thus, the statistical analysis highlights the thermal instability of the dice-bilayer, disintegrating at $700\,\text{K}$, versus the remarkable stability of the star-bilayer, whose melting temperature is enhanced to around $1200\,\text{K}$.

% Before the structural transition, the over-constrained dice lattice supports extensive self-stress states, enforcing rigidity and generating an incommensurate Moir\'{e} pattern on the platinum substrate. This produces localized self-stress fluctuations that trap elastic energy, ultimately leading to thermal instability. 
% After the transition, the star lattice balances constraints and degrees of freedom at the isostatic point, aligning with the substrate. This yields a commensurate Moir\'{e} pattern and a sub-extensive number of self-stress states more evenly distributed in space. As a result, elastic energy is less concentrated in bonds, markedly enhancing thermal resilience. 

\textit{Discussion---}We show that the exceptional thermal stability of platinum oxides originates from the rigidity of elastic networks governed by chemical constraints. In the dice lattice, over-constrained bonds induce localized self-stress and incommensurate Moir\'{e} patterns, leading to thermal instability. After transition, the isostatic star lattice balances constraints with degrees of freedom, aligns commensurately with the substrate, and yields delocalized self-stress states that enhance resilience. Our framework links thermal robustness to the topological index of self-stress, and extends naturally to other compounds~\cite{kim2006metalorganic,longo1971structure,moon2008dimensionality,okamoto2018transition,doennig2013massive,rawl2017magnetic,sumi2005mocvd}, elastic meta-surfaces~\cite{Ma2014, Li2016, Assouar2018, Oudich2010}, active frames~\cite{Zhou2022NC, Xu2023, WangMa2025}, and bio-material networks~\cite{ZhouHaiJun2005PRL}, offering a route to design materials with improved stability across scales.

%We demonstrate that the exceptional thermal stability of platinum oxides arises from the mechanical rigidity of elastic networks modeled by microscopic chemical constraints. In the initial dice lattice, over-constrained bonding supports extensive self-stress states, generating an incommensurate Moir\'e pattern on the platinum substrate and trapping elastic energy in localized self-stress fluctuations that ultimately lead to thermal instability. After the transition, the isostatic star lattice balances constraints with degrees of freedom and aligns commensurately with the substrate, yielding a sub-extensive set of more uniformly distributed self-stress states. This delocalized elastic energy markedly enhances thermal resilience.

%Our work establishes a universal framework for assessing the thermal robustness of chemical compounds, based solely on the topological index of self-stress states in network models. This approach generalizes readily to other chemical compounds~\cite{kim2006metalorganic,longo1971structure,moon2008dimensionality,okamoto2018transition,doennig2013massive,rawl2017magnetic,sumi2005mocvd}, elastic meta-surface sheets~\cite{Ma2014, Li2016, Assouar2018, Oudich2010}, active mechanical frames~\cite{Zhou2022NC, Xu2023, WangMa2025}, and bio-material networks~\cite{ZhouHaiJun2005PRL}, opening avenues for designing materials with enhanced thermal and mechanical stability across multiple length scales.

\textit{Acknowledgment---}D. Z. and F. M. would like to thank the insightful discussions with Yan-Wei Li and Peng Liu. This work is supported by National Natural Science Foundation of China Grant No. 12374157.

\section{Supplementary Materials}

\section{Stretching-induced force and constraint}

In this section, we derive the restoring forces arising from stretching elastic energy. We further derive the linearized constraint that leads to the compatibility matrix, which reflects the topological index of the state of self stress via the rank-nullity theorem. 

Consider a point mass $i$ in the elastic network, whose equilibrium position in three-dimensional space is $\bm{r}_i^{(0)}$. Elastic deformations and thermal fluctuations displace the site from equilibrium: $\bm{r}_i = \bm{r}_i^{(0)} + \bm{u}_i$, where $\bm{u}_i$ denotes the displacement vector of site $i$.

Let $\langle i,j\rangle$ denote the nearest-neighbor bond connecting sites $i$ and $j$. The undeformed bond vector is defined as $\bm{r}_{ij}^{(0)} = \bm{r}_j^{(0)} - \bm{r}_i^{(0)}$, with the corresponding unit vector
\begin{equation}
\hat{\bm{n}}_{ij} = {\bm{r}_{ij}^{(0)}}/{\ell_{\langle i,j\rangle}^{(0)}},
\end{equation}
where we define the rest length of bond $\langle i,j\rangle$ as $\ell_{\langle i,j\rangle}^{(0)}=|\bm{r}_{ij}^{(0)}|$. The deformed bond vector is given by $\bm{r}_{ij} = \bm{r}_{ij}^{(0)} + \bm{u}_{ij}$, $\bm{u}_{ij} = \bm{u}_j - \bm{u}_i$, where $\bm{u}_{ij}$ represents the relative displacement between the two sites. The edge length of the corresponding bond is defined as $\ell_{\langle i,j\rangle}=|\bm{r}_{ij}|$. Thus, the longitudinal stretching of bond $\langle i,j\rangle$ is defined as $\delta\ell_{\langle i,j\rangle} = \ell_{\langle i,j\rangle} - \ell_{\langle i,j\rangle}^{(0)}$. 

Within the framework of linear elasticity, we expand $\delta\ell_{ij}$ to first order in $\bm{u}_{ij}$:
\begin{equation}\label{11}
\delta\ell_{\langle i,j\rangle} = \hat{\bm{n}}_{ij} \cdot \bm{u}_{ij} + \mathcal{O}(\bm{u}_{ij}^2).
\end{equation}
The stretching force on site $i$ due to bond $\langle i,j\rangle$ is then
\begin{equation}
\bm{F}_{i,\langle i,j\rangle} = K_{ij}\,\mathbf{P}_{\langle i,j\rangle}\,\bm{u}_{ij},
\label{eq:stretching_force}
\end{equation}
where $K_{ij}$ is the stretching stiffness of the bond, and $\mathbf{P}_{\langle i,j\rangle}$ is the projection operator along the bond direction:
\begin{equation}
\mathbf{P}_{\langle i,j\rangle} = \hat{\bm{n}}_{ij}\,\hat{\bm{n}}_{ij}^{\mathsf{T}}.
\label{eq:projection_operator}
\end{equation}
Here $\hat{\bm{n}}_{ij}$ is treated as a $3\times1$ column vector, $\mathsf{T}$ denotes the transpose, and $\hat{\bm{n}}_{ij}\hat{\bm{n}}_{ij}^{\mathsf{T}}$ yields a $3\times3$ projection matrix. In the numerical simulations, we set all bond stretching stiffness to a uniform value, $K_{ij}=k$, for simplicity.

The \textit{stretching-induced compatibility matrix} $\mathbf{C}^{(s)}$ encodes the linear mapping from particle displacements to bond elongations within the harmonic approximation. For a network of $N$ particles connected by $N_{\rm nn}$ stretching bonds, we define: (i) The displacement vector $\bm{u} = (\bm{u}_1, \bm{u}_2, \dots, \bm{u}_N)^\mathsf{T} \in \mathbb{R}^{Nd}$, where $\bm{u}_i \in \mathbb{R}^3$ is the displacement of particle $i$. (ii) The bond-elongation vector $\delta\bm{\ell}^{(\rm nn)} = (\delta\ell^{(\rm nn)}_1, \delta\ell^{(\rm nn)}_2, \dots, \delta\ell^{(\rm nn)}_{N_{\rm nn}})^\mathsf{T} \in \mathbb{R}^{N_{\rm nn}}$, where each $\delta\ell_\alpha$ corresponds to the longitudinal stretch of a specific bond $\alpha \equiv \langle i,j\rangle$.

The compatibility matrix $\mathbf{C}^{(s)} \in \mathbb{R}^{N_{\rm nn} \times Nd}$ is then defined by the linear relation
\begin{equation}
\delta\bm{\ell}^{(\rm nn)} = \mathbf{C}^{(s)} \bm{u},
\label{eq:compatibility_relation}
\end{equation}
which holds to first order in the displacements. From the geometric expression $\delta\ell_{\langle i,j\rangle} = \hat{\bm{n}}_{ij} \cdot \bm{u}_{ij} + \mathcal{O}(\bm{u}^2)$ with $\bm{u}_{ij} = \bm{u}_j - \bm{u}_i$, the matrix elements of $\mathbf{C}^{(s)}$ follow directly. For a bond $\alpha = \langle i,j\rangle$, the only nonzero blocks in row $\alpha$ are
\begin{equation}
\mathbf{C}^{(s)}_{\alpha,i} = -\hat{\bm{n}}_{ij}^{\mathsf{T}}, \qquad
\mathbf{C}^{(s)}_{\alpha,j} = +\hat{\bm{n}}_{ij}^{\mathsf{T}},
\label{eq:compatibility_matrix_elements}
\end{equation}
where $\hat{\bm{n}}_{ij} \in \mathbb{R}^3$ is the unit vector along the bond in the reference configuration. All other entries $\mathbf{C}_{\alpha,k}^{(s)}$ for $k \neq i,j$ are zero. Thus, each row of $\mathbf{C}^{(s)}$ contains exactly two nonzero blocks of size $1 \times 3$, reflecting the fact that a bond stretch depends only on the relative displacement of its two endpoints.
%where $\langle i,j\rangle$ labels the bond. 

%The stretching contribution to the dynamical matrix for node $i$ follows as
%\begin{equation}
%\mathbf{D}_{ij}^{(s)} =
%\delta_{ij}\Bigl(\sum_{k} K_{ik}\,\mathbf{P}_{\langle i,k\rangle}\Bigr)
%- K_{ij}\,\mathbf{P}_{\langle i,j\rangle},
%\label{eq:dynamical_matrix_stretching}
%\end{equation}
%where the first term accounts for the restoring force from all bonds connected to site $i$, and the second term couples sites $i$ and $j$. 

\section{Bending-induced force and constraint}

In this section, we derive the restoring forces arising from the bending energy of elastic hinges. Based on the linearized restoring forces, we derive the compatibility matrix that stems from the bending constraints. Consider three neighboring sites $i$, $j$, $k$ in the network, where bonds $\langle i,j\rangle$ and $\langle j,k\rangle$ meet at vertex $j$. The angle $\angle ijk$, denoted by $\theta_{ijk}$, undergoes deformation due to displacements of the nodes, leading to bending elastic forces.

Let the equilibrium positions be $\bm{r}_i^{(0)}$, $\bm{r}_j^{(0)}$, $\bm{r}_k^{(0)}\in\mathbb{R}^3$, and the corresponding displacements be $\bm{u}_i$, $\bm{u}_j$, $\bm{u}_k$. The instantaneous positions are $\bm{r}_\alpha = \bm{r}_\alpha^{(0)} + \bm{u}_\alpha$, for $ \alpha = i,j,k$. The equilibrium angle $\Theta_{ijk}$ satisfies
\begin{eqnarray}
\cos\Theta_{ijk} = 
{\bm{r}_{ji}^{(0)}\cdot\bm{r}_{jk}^{(0)}}/{\ell_{ji}^{(0)}\,\ell_{jk}^{(0)}},
\end{eqnarray}
while the instantaneous angle $\theta_{ijk}$ follows from
\begin{eqnarray}
\cos\theta_{ijk} = 
{\bm{r}_{ji}\cdot\bm{r}_{jk}}/{\ell_{ji} \ell_{jk}}.
\end{eqnarray}
The bending energy associated with hinge $\langle i,j,k\rangle$ is
\begin{equation}
U_{\langle i,j,k\rangle} = \frac{1}{2}\,\kappa_{ijk}
\bigl(\theta_{ijk}-\Theta_{ijk}\bigr)^2,
\label{eq:bending_energy}
\end{equation}
where $\kappa_{ijk}$ is the bending stiffness.

Expanding the lengths to first order in displacements gives
\begin{align}
\ell_{ji} &\approx \ell_{ji}^{(0)} + 
{\bm{r}_{ji}^{(0)}\cdot\bm{u}_{ji}}/{\ell_{ji}^{(0)}}, \\
\ell_{jk} &\approx \ell_{jk}^{(0)} + 
{\bm{r}_{jk}^{(0)}\cdot\bm{u}_{jk}}/{\ell_{jk}^{(0)}}.
\end{align}
The dot product expands as
\begin{equation}
\bm{r}_{ji}\cdot\bm{r}_{jk} \approx 
\bm{r}_{ji}^{(0)}\cdot\bm{r}_{jk}^{(0)}
+ \bm{r}_{ji}^{(0)}\cdot\bm{u}_{jk}
+ \bm{r}_{jk}^{(0)}\cdot\bm{u}_{ji}.
\end{equation}
For small angular deviations, we use
\begin{equation}
\cos\theta_{ijk} - \cos\Theta_{ijk}
\approx -\bigl(\theta_{ijk}-\Theta_{ijk}\bigr)\,
\sin\Theta_{ijk}.
\end{equation}
After a straightforward but careful Taylor expansion, one obtains
\begin{equation}
\theta_{ijk}-\Theta_{ijk}
\approx 
- \frac{\bm{u}_{ji}\cdot\bigl(\mathbf{P}_{\langle j,i\rangle}^\perp\hat{\bm{n}}_{jk}\bigr)}
      {\ell_{ji}^{(0)}\sin\Theta_{ijk}}
- \frac{\bm{u}_{jk}\cdot\bigl(\mathbf{P}_{\langle j,k\rangle}^\perp\hat{\bm{n}}_{ji}\bigr)}
      {\ell_{jk}^{(0)}\sin\Theta_{ijk}},
\label{eq:angle_deviation}
\end{equation}
where $\mathbf{P}_{\langle j,i\rangle}^\perp$ is the transverse projection operator
\begin{equation}
\mathbf{P}_{\langle j,i\rangle}^\perp = 
\mathbf{I}_3 - \hat{\bm{n}}_{ji}\hat{\bm{n}}_{ji}^{\mathsf{T}}.
\end{equation}
From SI.~Eq.~\eqref{eq:angle_deviation},
\begin{equation}
\frac{\partial\theta_{ijk}}{\partial\bm{u}_j}
\approx 
\frac{\mathbf{P}_{\langle j,i\rangle}^\perp\hat{\bm{n}}_{jk}}
     {\ell_{ji}^{(0)}\sin\Theta_{ijk}}
+ \frac{\mathbf{P}_{\langle j,k\rangle}^\perp\hat{\bm{n}}_{ji}}
     {\ell_{jk}^{(0)}\sin\Theta_{ijk}}.
\label{eq:dtheta_du}
\end{equation}
The force on site $j$ due to hinge $\langle i,j,k\rangle$ is
\begin{equation}
\bm{F}_{j,\langle i,j,k\rangle} 
= -\frac{\partial U_{\langle i,j,k\rangle}}{\partial\bm{u}_j}.
\end{equation}
Consequently,
\begin{equation}
\bm{F}_{j,\langle i,j,k\rangle} 
= -\kappa_{ijk}\bigl(\theta_{ijk}-\Theta_{ijk}\bigr)\,
   \frac{\partial\theta_{ijk}}{\partial\bm{u}_j}.
\end{equation}
Substituting SI.~Eq.~\eqref{eq:angle_deviation} and SI.~Eq.~\eqref{eq:dtheta_du} yields the explicit linearized force:
\begin{multline}
\bm{F}_{j,\langle i,j,k\rangle} 
= \frac{\kappa_{ijk}}{\sin^2\Theta_{ijk}}
\Biggl(
\frac{\mathbf{P}_{\langle j,i\rangle}^\perp\hat{\bm{n}}_{jk}}{\ell_{ji}^{(0)}}
+ \frac{\mathbf{P}_{\langle j,k\rangle}^\perp\hat{\bm{n}}_{ji}}{\ell_{jk}^{(0)}}
\Biggr) \\
\cdot
\Biggl[
\frac{\bm{u}_{ji}\cdot\bigl(\mathbf{P}_{\langle j,i\rangle}^\perp\hat{\bm{n}}_{jk}\bigr)}
     {\ell_{ji}^{(0)}}
+ \frac{\bm{u}_{jk}\cdot\bigl(\mathbf{P}_{\langle j,k\rangle}^\perp\hat{\bm{n}}_{ji}\bigr)}
     {\ell_{jk}^{(0)}}
\Biggr].
\label{eq:bending_force_central}
\end{multline}

\section{Equivalence Between Bending Stiffness and Next-Nearest-Neighbor Interactions}

\begin{figure}[htbp]
\includegraphics[scale=0.7]{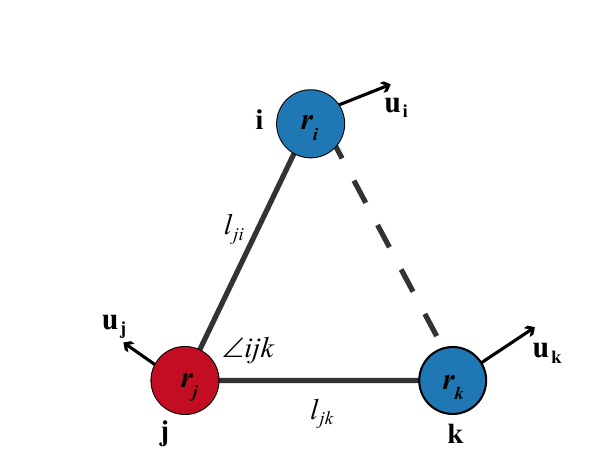}
\caption{Three adjacent sites in a spring-mass model.}\label{SIfig1}
\end{figure}

% This   These displacements perturb the hinge angle from $\Theta_{ijk}$ to $\theta_{ijk}$, where $|\theta_{ijk}-\Theta_{ijk}|\ll \Theta_{ijk}$. 

Within the linear elastic regime, and under the assumption that stretching stiffness is larger than bending-induced hinge stiffness, a local hinge stiffness can be equivalently mapped to an effective next-nearest-neighbor (NNN) stretching interaction~\cite{Lubensky2019PRL}. As illustrated in SI.~Fig.~\ref{SIfig1}, we consider three sites $i$, $j$, and $k$, with the central site $j$ bonded to its nearest neighbors $i$ and $k$. Their equilibrium positions are denoted $\bm{r}_i$, $\bm{r}_j$, and $\bm{r}_k$, respectively. The rest angle at the hinge, $\angle ijk = \Theta_{ijk}$, ensures that the bending energy vanishes in the reference configuration. The equilibrium geometry imposes the constraint
\begin{equation}
|\bm{r}_i-\bm{r}_j|^2 + |\bm{r}_k-\bm{r}_j|^2 - 2 |\bm{r}_i-\bm{r}_j| |\bm{r}_k-\bm{r}_j| \cos \Theta_{ijk} = |\bm{r}_k-\bm{r}_i|^2 .
\end{equation}

Applying small displacements $\bm{u}_i$, $\bm{u}_j$, and $\bm{u}_k$ to sites $i$, $j$, and $k$, the deformed positions satisfy
\begin{eqnarray}
& {} & |\bm{r}_i+\bm{u}_i-\bm{r}_j-\bm{u}_j|^2 + |\bm{r}_k+\bm{u}_k-\bm{r}_j-\bm{u}_j|^2 \nonumber \\
& {} & - 2 |\bm{r}_i+\bm{u}_i-\bm{r}_j-\bm{u}_j| |\bm{r}_k+\bm{u}_k-\bm{r}_j-\bm{u}_j| \cos \theta_{ijk} \nonumber \\
& {} & = |\bm{r}_k+\bm{u}_k-\bm{r}_i-\bm{u}_i|^2 .
\end{eqnarray}
The change of the bending angle from $\Theta_{ijk}$ to $\theta_{ijk}$ reproduces the results of SI.~Eq.~\eqref{eq:bending_energy}–\eqref{eq:bending_force_central}. Under the approximation that stretching stiffness greatly exceeds bending stiffness, stretching deformations are negligible compared to bending, yielding
\begin{equation}\label{4}
|\bm{u}_i-\bm{u}_j|, \; |\bm{u}_k-\bm{u}_j| \ll |\bm{u}_i-\bm{u}_k| \ll |\bm{r}_i-\bm{r}_k|.
\end{equation}
Thus, the perturbed configuration satisfies
\begin{eqnarray}
& {} & |\bm{r}_i+\bm{u}_i-\bm{r}_j-\bm{u}_j| \approx |\bm{r}_i-\bm{r}_j|, \nonumber \\
& {} & |\bm{r}_k+\bm{u}_k-\bm{r}_j-\bm{u}_j| \approx |\bm{r}_k-\bm{r}_j|.
\end{eqnarray}
Within these approximations, the distance between $i$ and $k$ becomes
\begin{eqnarray}\label{eq:NNNapprox}
& {} & |\bm{r}_i-\bm{r}_j|^2 + |\bm{r}_k-\bm{r}_j|^2 - 2 |\bm{r}_i-\bm{r}_j| |\bm{r}_k-\bm{r}_j| \cos \theta_{ijk} \nonumber \\
& {} & \approx \left|\bm{r}_i+\bm{u}_i - \bm{r}_k - \bm{u}_k \right|^2 .
\end{eqnarray}
Expanding both sides to leading order in $\Theta_{ijk}-\theta_{ijk}$ and $\bm{u}_i-\bm{u}_k$ yields
\begin{equation}
\Theta_{ijk}-\theta_{ijk} =
\frac{|\bm{r}_k - \bm{r}_i|}{|\bm{r}_i-\bm{r}_j| |\bm{r}_k-\bm{r}_j| \sin \Theta_{ijk}}\,
\hat{n}_{ik} \cdot (\bm{u}_k - \bm{u}_i),
\end{equation}
where $\hat{n}_{ik} = (\bm{r}_k-\bm{r}_i)/|\bm{r}_k-\bm{r}_i|$ is the unit vector from $i$ to $k$. Substituting into SI.~Eq.~\eqref{eq:bending_energy} with bending stiffness $\kappa_{ijk}$ gives
\begin{equation}
U_{\langle i, j, k\rangle}
= \frac{K_{ijk}}{2} \left(\delta \ell^{(\rm nnn)}_{ik}\right)^2,
\end{equation}
as if a virtual bond of stiffness $K_{ijk}$ directly connected $i$ and $k$, with
\begin{equation}
K_{ijk} = \kappa_{ijk}  \frac{|\bm{r}_k - \bm{r}_i|^2}{|\bm{r}_i-\bm{r}_j|^2 |\bm{r}_k-\bm{r}_j|^2 \sin^2 \Theta_{ijk}} ,
\end{equation}
and
\begin{equation}
\delta\ell^{(\rm nnn)}_{ik} = \hat{n}_{ik} \cdot (\bm{u}_k - \bm{u}_i).
\end{equation}

Thus, within linear elasticity, hinge bending is equivalent to a next-nearest-neighbor Hookean interaction between $i$ and $k$ (see Ref.~\cite{PhysRevLett.122.128001} for simulation-based validation). This approximation—replacing bending constraints with effective next-nearest-neighbor Hookean interactions—has become standard in both analytical and numerical studies of elastic networks \cite{Lubensky2019PRL, PhysRevLett.122.128001}. Following this methodology, we construct the compatibility matrix $\mathbf{C}^{(b)}$ by expressing bending constraints in terms of next-nearest-neighbor interactions.

\textit{Bending compatibility matrix.} The matrix $\mathbf{C}^{(b)}$ encodes the linear relation between particle displacements and changes in bending angles, equivalently represented as length changes in virtual next-nearest-neighbor bonds. As shown in Fig.~1 of the main text, each hinge angle is mapped to a corresponding next-nearest-neighbor bond. For a network with $N$ vertices and $N_{\rm nnn}$ such constraints, we define the bond-change vector $\delta\bm{\ell}^{(\rm nnn)} = (\delta\ell^{(\rm nnn)}_{1}, \delta\ell^{(\rm nnn)}_{2}, \dots, \delta\ell^{(\rm nnn)}_{N_{\rm nnn}})^{\mathsf{T}} \in \mathbb{R}^{N_{\rm nnn}}$, where each component $\delta\ell^{(\rm nnn)}_\alpha$ corresponds to the angular deviation of hinge that maps to the next-nearest-neighbor bond $\alpha \equiv \langle i,k\rangle$, from its equilibrium length that corresponds to the opening angle of $\Theta_{ijk}$. Within the harmonic approximation, $\mathbf{C}^{(b)} \in \mathbb{R}^{N_{\rm nnn} \times Nd}$ is defined by
\begin{equation}
\delta\bm{\ell}^{(\rm nnn)} = \mathbf{C}^{(b)} \bm{u},
\label{eq:bending_compatibility_relation}
\end{equation}
valid to first order in displacements $\bm{u} \in \mathbb{R}^{Nd}$. For a virtual next-nearest-neighbor constraint $\alpha = \langle i, k\rangle$, the only nonzero blocks in row $\alpha$ are
\begin{equation}
\mathbf{C}^{(b)}_{\alpha,i} = -\hat{\bm{n}}_{ik}^{\mathsf{T}}, \qquad
\mathbf{C}^{(b)}_{\alpha,k} = +\hat{\bm{n}}_{ik}^{\mathsf{T}},
\label{eq:compatibility_matrix_elements}
\end{equation}
with all other entries $\mathbf{C}_{\alpha,a}^{(b)}$ for $a \neq i,k$ equal to zero. Each row of $\mathbf{C}^{(b)}$ therefore contains exactly two nonzero $1 \times 3$ blocks, reflecting that a next-nearest-neighbor bond stretch depends only on the relative displacement of its two endpoints.

%The corresponding effective spring constant is
%\begin{equation}
%K =4\kappa \frac{|\bm{r}_2 - \bm{r}_1|^2}{S_{DOC}^2}.
%\end{equation}
%where
%\begin{equation}
%S_{DOC} = \frac{1}{2} | l_1 l_2 \sin \theta|
%\end{equation}
%is the area of the triangle $\Delta DOC$.

%The explicit form of $\mathbf{C}^{(b)}$ follows directly from the linearized geometric relation in ~\eqref{eq:angle_deviation}. For a hinge $\alpha = \langle i,j,k\rangle$ centered at vertex $j$, the non-zero $1 \times 3$ blocks in row $\alpha$ are:
%\begin{align}
%\mathbf{C}^{(b)}_{\alpha,i} &= 
%-\frac{\bigl(\mathbf{P}_{\langle j,i\rangle}^\perp \hat{\bm{n}}_{jk}\bigr)^{\!\mathsf{T}}}
%      {\ell_{ji}^{(0)}\sin\Theta_{ijk}},\\[4pt]
%\mathbf{C}^{(b)}_{\alpha,k} &= 
%-\frac{\bigl(\mathbf{P}_{\langle j,k\rangle}^\perp \hat{\bm{n}}_{ji}\bigr)^{\!\mathsf{T}}}
%      {\ell_{jk}^{(0)}\sin\Theta_{ijk}},\\[4pt]
%\mathbf{C}^{(b)}_{\alpha,j} &= 
%-\bigl(\mathbf{C}^{(b)}_{\alpha,i} + \mathbf{C}^{(b)}_{\alpha,k}\bigr),
%\label{eq:bending_compatibility_elements}
%\end{align}
%where $\mathbf{C}^{(b)}_{\alpha,p}$ denotes the $1\times 3$ block corresponding to vertex $p$ ($p=i,j,k$). All other blocks $\mathbf{C}^{(b)}_{\alpha,l}$ for $l \neq i,j,k$ vanish. Each row of $\mathbf{C}^{(b)}$ therefore contains exactly three non-zero blocks, reflecting the fact that the change in a hinge angle depends only on the displacements of its three constituent vertices.

\section{Rank--nullity theorem and the number of states of self-stress}

Having defined both stretching and bending compatibility matrices, we now combine them into the total compatibility matrix
\begin{equation}
\mathbf{C} = \begin{pmatrix} \mathbf{C}^{(s)} \\[2pt] \mathbf{C}^{(b)} \end{pmatrix}.
\end{equation}
Its dimensions are
\begin{equation}
\mathbf{C} \in \mathbb{R}^{N_{\mathrm{c}} \times Nd}, \qquad 
N_{\mathrm{c}} \equiv N_{\rm nn} + N_{\rm nnn},
\end{equation}
where $N_{\rm nn}$ is the number of stretching constraints (bonds), $N_{\rm nnn}$ the number of bending constraints (hinges), and $N_{\mathrm{c}}$ the total number of constraints. 

%Although $\mathbf{C}^{(b)}$ carries units of inverse length (due to the denominators $\ell^{(0)}$) while $\mathbf{C}^{(s)}$ is dimensionless, this difference does not affect the null space or rank of $\mathbf{C}$, because a uniform rescaling of rows does not change the linear-algebraic properties relevant here.

States of self-stress (SSS) are internal force distributions that satisfy force balance on every node without external loading. In the compatibility-matrix formalism, a state of self-stress corresponds to a vector $\bm{t} \in \mathbb{R}^{N_{\mathrm{c}}}$ of bond tensions and angular moments such that $\mathbf{C}^{\!\mathsf{T}}\bm{t} = \bm{0}$, i.e., $\bm{t}$ lies in the kernel (null space) of $\mathbf{C}^{\!\mathsf{T}}$. 
Thus the number of independent SSS is $N_{\mathrm{sss}} = \dim \ker \bigl(\mathbf{C}^{\!\mathsf{T}}\bigr) 
\equiv \operatorname{null}\bigl(\mathbf{C}^{\!\mathsf{T}}\bigr)$. For any matrix $\mathbf{A} \in \mathbb{R}^{m \times n}$, the rank--nullity theorem states $\operatorname{rank}(\mathbf{A}) + \operatorname{null}(\mathbf{A}) = n$, where $\operatorname{null}(\mathbf{A}) = \dim\ker(\mathbf{A})$. 
Applying this to $\mathbf{A} = \mathbf{C}^{\!\mathsf{T}} \in \mathbb{R}^{Nd \times N_{\mathrm{c}}}$ gives
\begin{equation}
\operatorname{rank}\bigl(\mathbf{C}^{\!\mathsf{T}}\bigr) 
+ \operatorname{null}\bigl(\mathbf{C}^{\!\mathsf{T}}\bigr) = N_{\mathrm{c}}.
\label{eq:rank_nullity_transpose}
\end{equation}
The rank of $\mathbf{C}^{\!\mathsf{T}}$ equals the rank of $\mathbf{C}$, which is at most $Nd$. For a rigid structure that has no zero-energy deformation modes (i.e., $\ker(\mathbf{C})$ is trivial), we have $\operatorname{rank}\bigl(\mathbf{C}^{\!\mathsf{T}}\bigr) = Nd$. Substituting this into SI.~Eq.\eqref{eq:rank_nullity_transpose} yields
\begin{equation}
N_{\mathrm{sss}} 
= \operatorname{null}\bigl(\mathbf{C}^{\!\mathsf{T}}\bigr)
= N_{\mathrm{c}} - Nd.
\label{eq:maxwell_index_3d}
\end{equation}
Equation~\eqref{eq:maxwell_index_3d} is the topological counting relation used in the main text to quantify the number of states of self-stress in over-constrained bilayer networks.

%Correspondingly, the rank of $\mathbf{C}$ relates through the rank-nullity theorem to the number of independent states of self-stress:

\section{Rescaling of the Newtonian equation of motion}

At atomic scales, physical quantities span extremely small magnitudes, complicating the direct numerical integration of the Langevin equation. For example, the atomic mass unit is $1\,\text{u}=1.66\times10^{-27}$\,kg, interatomic distances are of order $\mathcal{O}(10\,\text{\AA})$, and Boltzmann’s constant is $k_{\!B}=1.38\times10^{-23}$\,J/K. Such tiny numbers lead to extremely slow numerical evolution unless the equations are properly rescaled. To circumvent this issue, we rescale the Newtonian equation of motion so that the effective spring constant and damping coefficient both become unity, thereby accelerating the simulations while preserving the correct physics.

To illustrate the rescaling procedure, consider a single particle governed by a Langevin equation with a harmonic restoring force and thermal noise:
\begin{equation}
m\frac{\Delta^{2}\bm{u}}{\Delta t^{2}}
= -k\bm{u}
- \eta\frac{\Delta\bm{u}}{\Delta t}
+ \sqrt{\frac{2\eta k_{\!B}T}{\Delta t}}\;\bm{\xi}(t),
\label{eq:langevin_original}
\end{equation}
where $m$ is the particle mass, $\bm{u}$ the displacement from equilibrium, $k$ the spring stiffness, $\eta$ the damping coefficient, $T$ the temperature, $\Delta t$ the numerical time step, and $\bm{\xi}(t)$ a three‑dimensional white‑noise vector with $\langle\xi_{\mu}(t)\xi_{\nu}(t')\rangle=\delta_{\mu\nu}\delta_{tt'}$. 

We introduce dimensionless variables via $\Delta t = \alpha\Delta\tilde{t}$, $\bm{u} = \beta\tilde{\bm{u}}$, $m_{\rm O} = m_0\tilde{m}_{\rm O}$, $m_{\rm Pt} = m_0\tilde{m}_{\rm Pt}$, $T = T_{0}\,\tilde{T}$, with $m_0 = 16\text{u}$ (the mass of an oxygen atom) and $T_0=10^{7}$\,K. In this way, the dimensionless masses for oxygen and platinum atoms are $\tilde{m}_{\rm O}=1$ and $\tilde{m}_{\rm Pt} = 12.1875$, respectively. Substituting these into SI.~Eq.~\eqref{eq:langevin_original} and choosing
\[
\alpha = \sqrt{\frac{m_{\rm O}}{k}},\qquad
\beta = \sqrt{\frac{\alpha^{3}\eta k_{\!B}T_{0}}{m_{\rm O}}},
\]
the equation transforms to
\begin{equation}
\tilde{m}\,\frac{\Delta^{2}\tilde{\bm{u}}}{\Delta\tilde{t}^{2}}
= -\tilde{\bm{u}}
- \frac{\Delta\tilde{\bm{u}}}{\Delta\tilde{t}}
+ \sqrt{\frac{2\tilde{T}}{\Delta\tilde{t}}}\;\bm{\xi},
\label{eq:langevin_rescaled}
\end{equation}
in which the effective spring constant and damping coefficient are both unity.

For a typical spring stiffness $k\approx100$\,N/m, we obtain $\alpha = 1.628\times10^{-14}\;\text{s}$, $\beta = 1.170\times10^{-9}\;\text{m}$, and the required physical damping coefficient becomes $\eta = \sqrt{m_{\rm O}k}\approx 1.628\times10^{-12}\;\text{kg/s}$. This value is consistent with the order of magnitude expected from Stokes’ law, $\eta\sim6\pi\gamma a$, where $a$ is the particle radius and $\gamma\sim10^{-4}$\,kg$\cdot$m$^{-1}\cdot$s$^{-1}$ the viscosity of $\text{NO}_{2}$. With the above choice of $\alpha$, $\beta$, and $\eta$, the rescaled equation \eqref{eq:langevin_rescaled} is free of extremely small numerical prefactors and can be integrated efficiently while retaining the correct statistical and dynamical properties.

\begin{figure}[htbp]
	\includegraphics[scale=0.42]{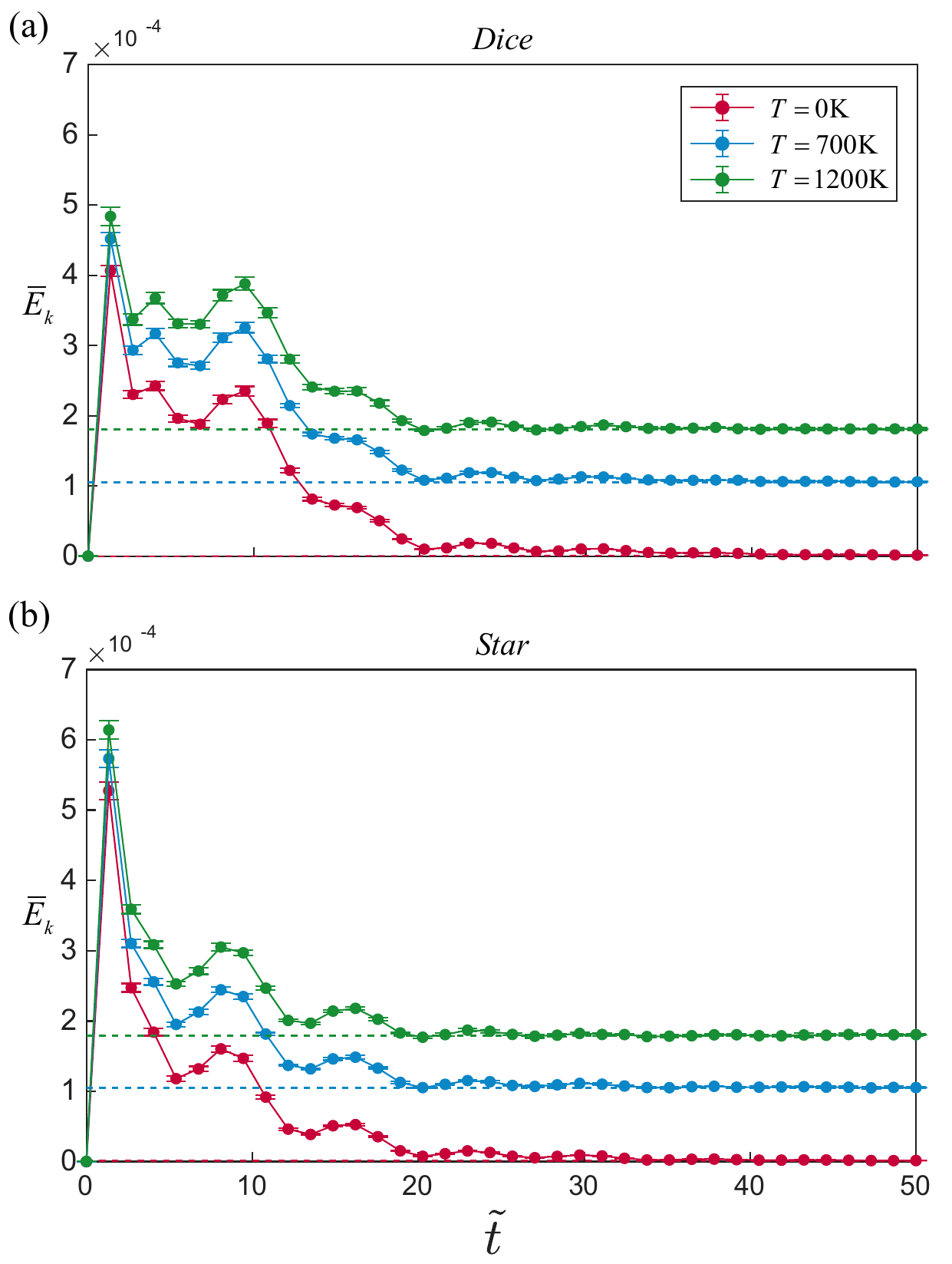}
	\caption{Time evolution of the average dimensionless molecular kinetic energy for (a) the dice-bilayer and (b) the star-bilayer during relaxation at $0\, \text{K}$ (red curve), $700\, \text{K}$ (blue curve), and $1200\, \text{K}$ (green curve).}\label{SIfig2}
\end{figure}

\section{Numerical simulations}

Here, we present numerical details on the mechanical and thermal dynamics of the platinum oxide bilayers.

% Bilayer structures after mechanical relaxation. Spatial profiles of elastic energy in the dice (a) and star (b) lattices at 0K, with colors on nearest- and next-nearest-neighbor edges indicating stretching and bending contributions. Height distribution of platinum atoms in dice (c) and star (d) lattices at 0K, where $ h_{Pt}^{(0)}$ denotes the height of platinum atoms in the initial state.

\begin{figure}[htbp]
	\includegraphics[scale=0.41]{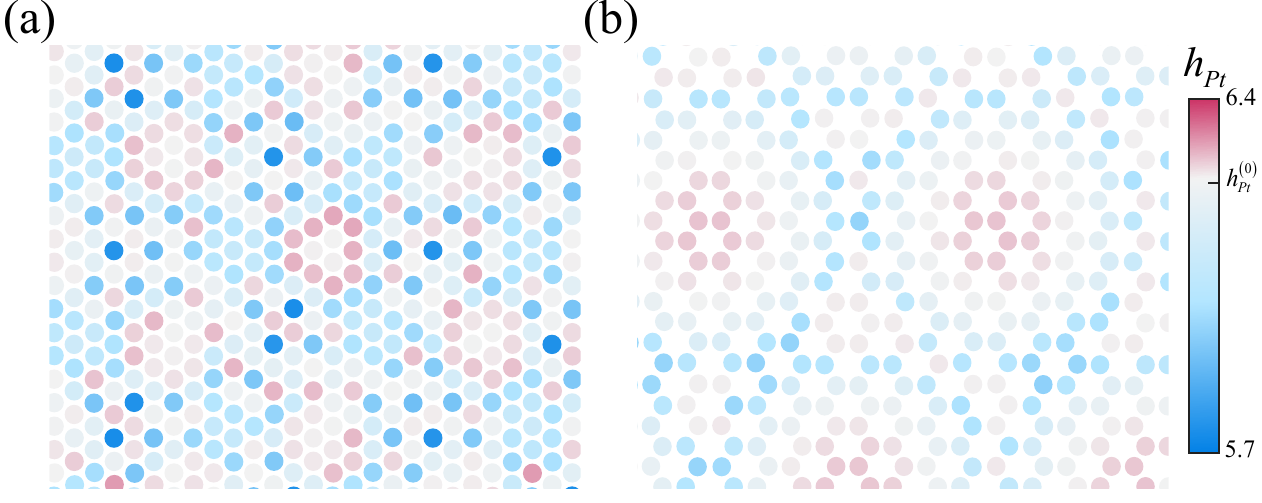}
	\caption{The height variations of platinum atoms in the dice and star platinum oxide lattices after mechanical relaxation at $0\,\text{K}$ for (a) dice lattice and (b) star lattice, respectively. White color corresponds to the height of platinum atom before relaxation. }\label{SIfig4} 
\end{figure}

\begin{figure}[htbp]
	\includegraphics[scale=0.41]{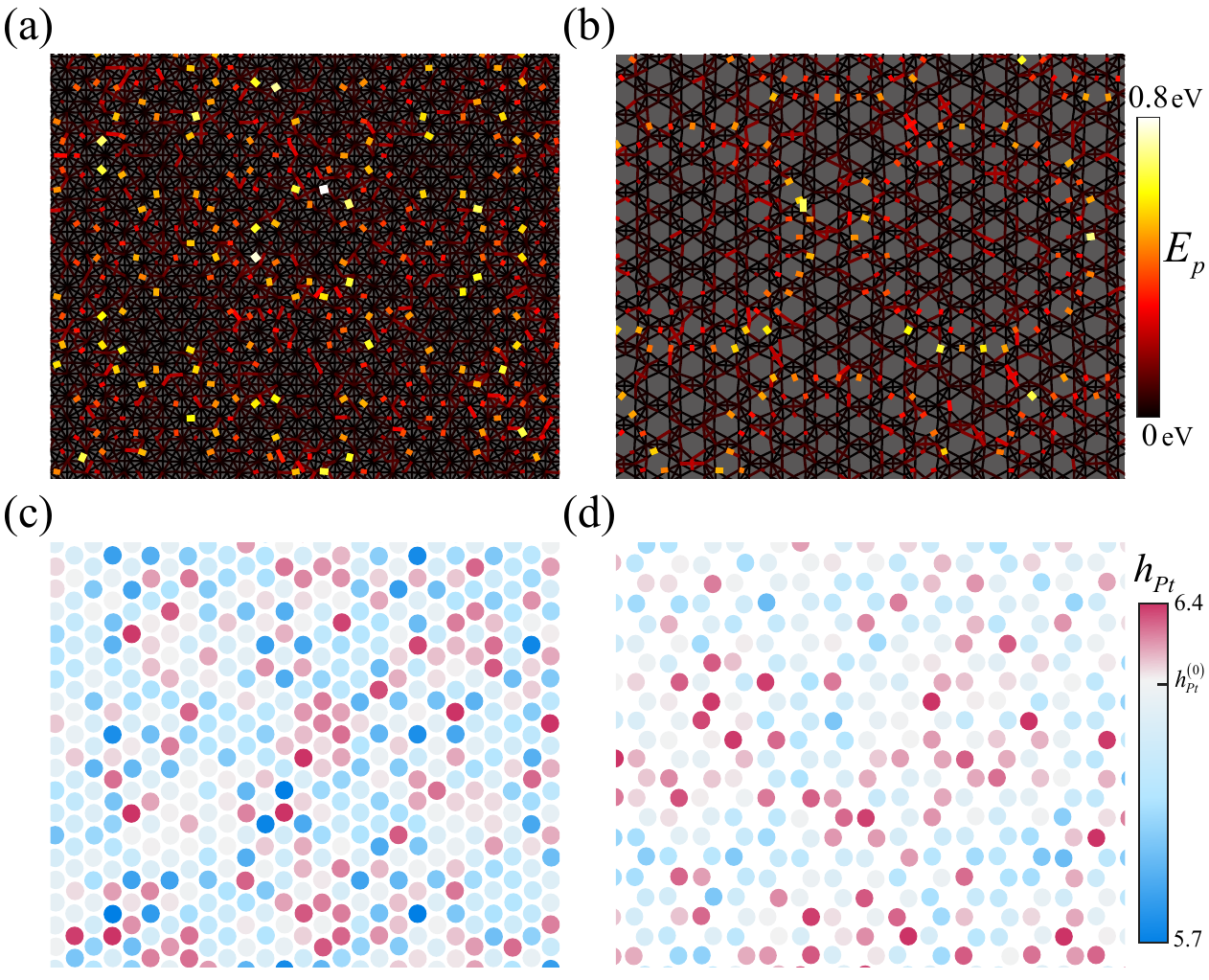}
	\caption{Spatial profiles of elastic energy in the (a) dice bilayer and (b) star bilayer at thermal equilibrium ($700\,\text{K}$). In (a,b) the color and width of each edge represent the magnitude of elastic energy stored in the corresponding constraint; nearest-neighbor edges show stretching contributions, next-nearest-neighbor edges show bending contributions. Height distributions of platinum atoms in the (c) dice and (d) star lattices at $700\,\text{K}$, where $h_{\rm Pt}^{(0)}$ denotes the height in the initial (pre-equilibrium) state.}\label{SIfig5}
\end{figure}

\subsection{Mechanical and thermal equilibrium of bilayer structures}

In the bilayer structures shown in Figs.~3(a,b) and Fig.~4 of the main text, the dice and star lattices contain $35 \times 35$ and $20 \times 20$ unit cells, respectively. These sizes are chosen so that the two systems have equal in-plane area. The underlying triangular platinum substrate is made sufficiently large to fully cover the overlying oxide layer. Open boundary conditions are applied to the \textit{four side edges} of the bilayer system. The top layer of platinum atoms is free to evolve dynamically, while the subsurface layer is held fixed to represent the rigid bulk substrate.

\begin{figure}[htbp]
\includegraphics[scale=0.41]{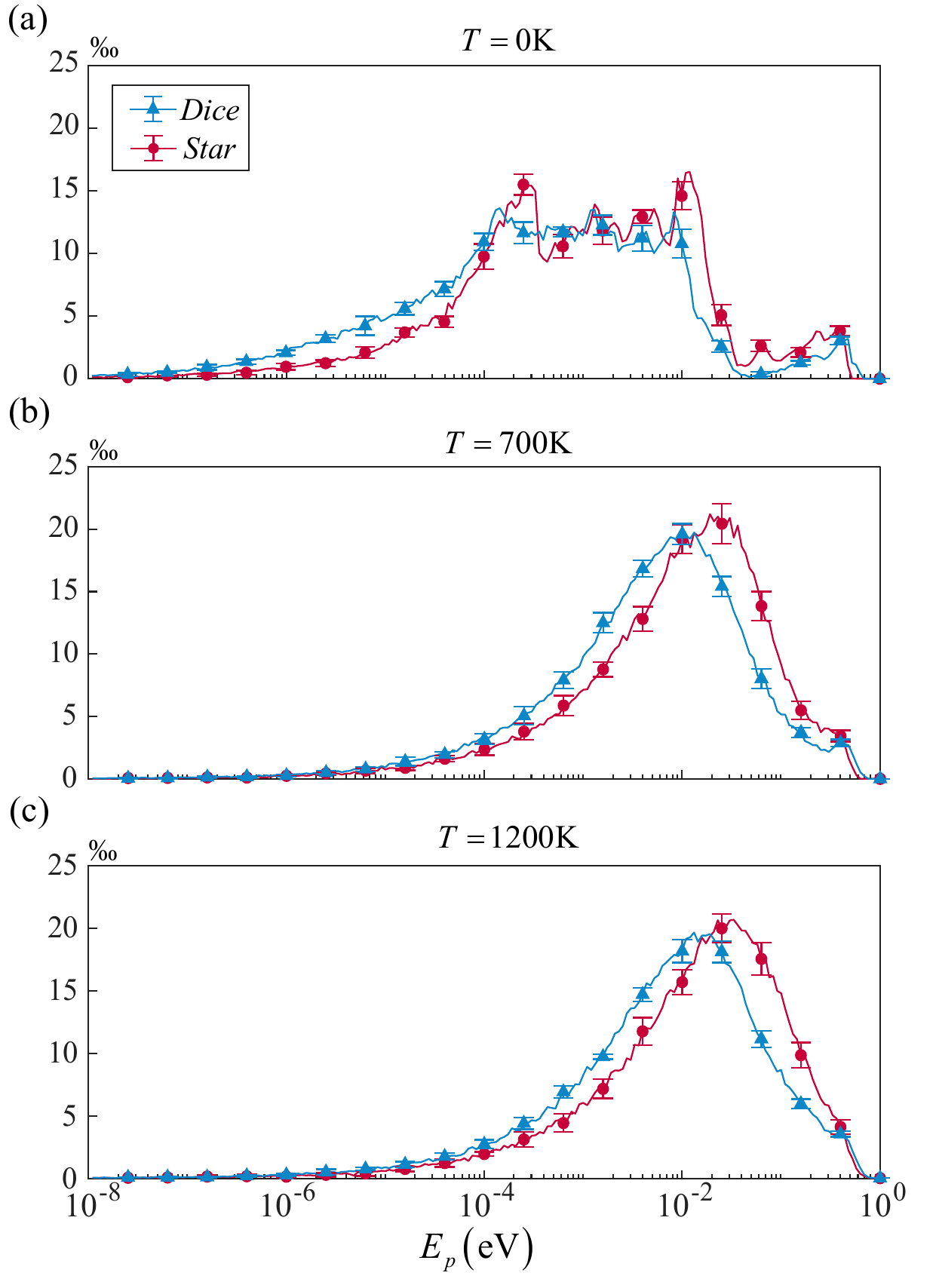}
\caption{Distributions of elastic energy per constraint for the dice (blue curves) and star (red curves) bilayers at (a) $0\ \text{K}$, (b) $700\ \text{K}$, and (c) $1200\ \text{K}$. The horizontal axis shows $\log_{10}(E/\text{eV})$, where $E$ is the elastic energy per constraint. Data points are placed at the centers of uniformly spaced intervals in $\log_{10}E$ with $\Delta(\log_{10}E)=0.04$; error bars indicate the standard deviation within each interval. The vertical axis gives the normalized probability density scaled in units of $10^{-3}$. Triangles and circles mark the mean values for the dice and star bilayers, respectively.}\label{SIfig6}
\end{figure}

We numerically integrate the dimensionless Langevin equation (SI.~Eq.~\eqref{eq:langevin_rescaled}) using a fourth-order Runge-Kutta method. The dimensionless temperatures $\tilde{T}$ correspond to $0$~K, $700$~K, and $1200$~K, and are set to $\tilde{T}=0$, $7 \times 10^{-5}$, and $1.2 \times 10^{-4}$, respectively. The integration time step is $\Delta\tilde{t}=10^{-3}$.

No measurements are taken until the system reaches either mechanical equilibrium (for $T=0$~K, when the elastic energy is minimized) or thermal equilibrium (for $T=700$~K and $1200$~K, when equipartition is established). Equilibrium is monitored by tracking the mean kinetic energy per particle, as shown in SI.~Fig.~\ref{SIfig2}. The horizontal dashed lines indicate the expected equilibrium values $\langle E\rangle = k_{B}T/2$ for each temperature. The total simulation time is $\tilde{t}_{\max}=200$. SI.~Fig.~\ref{SIfig2} shows that the average kinetic energy already stabilises well before $\tilde{t}{\text{criteria}}=50$, confirming that equilibrium is reached much earlier than $\tilde{t}_{\max}$: for $T=0$~K (red curve) the kinetic energy vanishes, while for $700$~K (blue) and $1200$~K (green) it saturates at $k_{B}\times700\,\text{K}/2$ and $k_{B}\times1200\,\text{K}/2$, respectively.

% The bilayer structures are relaxed to equilibrium at 0K, 700K and 1200K, respectively. 

\subsection{Additional simulation results of the bilayer structures after equilibrium at $0,\text{K}$ and $700,\text{K}$}

In addition to the elastic energy stored in each constraint after mechanical relaxation at $0\,$K, as presented in Fig.~3 of the main text, we also exhibit the height variation of the platinum atoms in the dice and star platinum oxide lattices, which further reflect the quasicrystalline and superlattice structures of the bilayers. As shown in SI.~Figs.~\ref{SIfig4}(a) and (b), after mechanical relaxation, the height of the platinum atoms in the lattices of platinum oxides are recorded after mechanical relaxation. The dice bilayer structure exhibits a quasicrystalline and highly fluctuating pattern for the platinum atom height, in accordance with the quasicrystalline profile of the elastic energy stored in the elastic frame, whereas for the star lattice, the crystalline pattern of platinum height variations are in line with the crystalline pattern stored in the elastic energy of the dice lattice after mechanical relaxation, as shown by Fig.~3 in the main text.

We further compute the spatial profiles of the elastic energy stored in each constraint and the platinum atom height across both bilayer structures, at the temperature of $700\,\text{K}$, namely the melting temperature of the dice bilayer. As shown in SI.~Figs.~\ref{SIfig5} (a) and (b), while we are considering the finite-temperature case as compared to the zero-temperature limit in Fig.~3, the pattern of the elastic energy stored in each constraint only exhibits an overall increased magnitude, while the quasicrystalline and superlattice patterns in the dice-bilayer and star-bilayer remain unchanged. 

Moreover, comparing SI.~Figs.~\ref{SIfig5} (c,d) to SI.~Figs.~\ref{SIfig4}, it is shown that the fluctuations in the height of the platinum atoms in the platinum oxides fluctuate more randomly due to the finite-temperature effect. However, the spatial profile of the platinum height in the star lattice still exhibit a crystalline structure comparing to the highly fluctuating pattern in the dice lattice, reflecting the quasicrystalline and superlattice nature of the dice and star bilayer, respectively. The distinct height profiles of the two lattices also govern their thermal responses: the periodic height modulations of the star lattice can distribute thermally induced elastic energy over extended regions, while the localized, fluctuating height variations in the dice lattice concentrate deformation on short scales, which critically undermines its thermal stability.

\subsection{Complete distributions of stretching and bending energies per bond at $0\,\text{K}$, $700\,\text{K}$, and $1200\,\text{K}$}

The full distributions of the elastic energy per bond (stretching and bending) in the dice and star bilayers at $0\,\text{K}$, $700\,\text{K}$, and $1200\,\text{K}$ are presented in SI.~Fig.~\ref{SIfig6}. A magnified view of the high-energy region above $0.5\,\text{eV}$ is shown in Fig.~4 of the main text.

%\bibliography{DiZhou_BIB}

\begin{thebibliography}{65}%
\makeatletter
\providecommand \@ifxundefined [1]{%
 \@ifx{#1\undefined}
}%
\providecommand \@ifnum [1]{%
 \ifnum #1\expandafter \@firstoftwo
 \else \expandafter \@secondoftwo
 \fi
}%
\providecommand \@ifx [1]{%
 \ifx #1\expandafter \@firstoftwo
 \else \expandafter \@secondoftwo
 \fi
}%
\providecommand \natexlab [1]{#1}%
\providecommand \enquote  [1]{``#1''}%
\providecommand \bibnamefont  [1]{#1}%
\providecommand \bibfnamefont [1]{#1}%
\providecommand \citenamefont [1]{#1}%
\providecommand \href@noop [0]{\@secondoftwo}%
\providecommand \href [0]{\begingroup \@sanitize@url \@href}%
\providecommand \@href[1]{\@@startlink{#1}\@@href}%
\providecommand \@@href[1]{\endgroup#1\@@endlink}%
\providecommand \@sanitize@url [0]{\catcode `\\12\catcode `\$12\catcode `\&12\catcode `\#12\catcode `\^12\catcode `\_12\catcode `\%12\relax}%
\providecommand \@@startlink[1]{}%
\providecommand \@@endlink[0]{}%
\providecommand \url  [0]{\begingroup\@sanitize@url \@url }%
\providecommand \@url [1]{\endgroup\@href {#1}{\urlprefix }}%
\providecommand \urlprefix  [0]{URL }%
\providecommand \Eprint [0]{\href }%
\providecommand \doibase [0]{http://dx.doi.org/}%
\providecommand \selectlanguage [0]{\@gobble}%
\providecommand \bibinfo  [0]{\@secondoftwo}%
\providecommand \bibfield  [0]{\@secondoftwo}%
\providecommand \translation [1]{[#1]}%
\providecommand \BibitemOpen [0]{}%
\providecommand \bibitemStop [0]{}%
\providecommand \bibitemNoStop [0]{.\EOS\space}%
\providecommand \EOS [0]{\spacefactor3000\relax}%
\providecommand \BibitemShut  [1]{\csname bibitem#1\endcsname}%
\let\auto@bib@innerbib\@empty
%</preamble>
\bibitem [{\citenamefont {Park}\ \emph {et~al.}(2000)\citenamefont {Park}, \citenamefont {Vohs},\ and\ \citenamefont {Gorte}}]{park2000direct}%
  \BibitemOpen
  \bibfield  {author} {\bibinfo {author} {\bibfnamefont {S.}~\bibnamefont {Park}}, \bibinfo {author} {\bibfnamefont {J.~M.}\ \bibnamefont {Vohs}}, \ and\ \bibinfo {author} {\bibfnamefont {R.~J.}\ \bibnamefont {Gorte}},\ }\href@noop {} {\bibfield  {journal} {\bibinfo  {journal} {Nature}\ }\textbf {\bibinfo {volume} {404}},\ \bibinfo {pages} {265} (\bibinfo {year} {2000})}\BibitemShut {NoStop}%
\bibitem [{\citenamefont {Van~Spronsen}\ \emph {et~al.}(2017)\citenamefont {Van~Spronsen}, \citenamefont {Frenken},\ and\ \citenamefont {Groot}}]{van2017observing}%
  \BibitemOpen
  \bibfield  {author} {\bibinfo {author} {\bibfnamefont {M.~A.}\ \bibnamefont {Van~Spronsen}}, \bibinfo {author} {\bibfnamefont {J.~W.}\ \bibnamefont {Frenken}}, \ and\ \bibinfo {author} {\bibfnamefont {I.~M.}\ \bibnamefont {Groot}},\ }\href@noop {} {\bibfield  {journal} {\bibinfo  {journal} {Nature communications}\ }\textbf {\bibinfo {volume} {8}},\ \bibinfo {pages} {429} (\bibinfo {year} {2017})}\BibitemShut {NoStop}%
\bibitem [{\citenamefont {Wilson}\ and\ \citenamefont {Lippard}(2014)}]{wilson2014synthetic}%
  \BibitemOpen
  \bibfield  {author} {\bibinfo {author} {\bibfnamefont {J.~J.}\ \bibnamefont {Wilson}}\ and\ \bibinfo {author} {\bibfnamefont {S.~J.}\ \bibnamefont {Lippard}},\ }\href@noop {} {\bibfield  {journal} {\bibinfo  {journal} {Chemical reviews}\ }\textbf {\bibinfo {volume} {114}},\ \bibinfo {pages} {4470} (\bibinfo {year} {2014})}\BibitemShut {NoStop}%
\bibitem [{\citenamefont {Steininger}\ \emph {et~al.}(1982)\citenamefont {Steininger}, \citenamefont {Lehwald},\ and\ \citenamefont {Ibach}}]{steininger1982adsorption}%
  \BibitemOpen
  \bibfield  {author} {\bibinfo {author} {\bibfnamefont {H.}~\bibnamefont {Steininger}}, \bibinfo {author} {\bibfnamefont {S.}~\bibnamefont {Lehwald}}, \ and\ \bibinfo {author} {\bibfnamefont {H.}~\bibnamefont {Ibach}},\ }\href@noop {} {\bibfield  {journal} {\bibinfo  {journal} {Surface Science}\ }\textbf {\bibinfo {volume} {123}},\ \bibinfo {pages} {1} (\bibinfo {year} {1982})}\BibitemShut {NoStop}%
\bibitem [{\citenamefont {Devarajan}\ \emph {et~al.}(2008)\citenamefont {Devarajan}, \citenamefont {Hinojosa~Jr},\ and\ \citenamefont {Weaver}}]{devarajan2008stm}%
  \BibitemOpen
  \bibfield  {author} {\bibinfo {author} {\bibfnamefont {S.~P.}\ \bibnamefont {Devarajan}}, \bibinfo {author} {\bibfnamefont {J.~A.}\ \bibnamefont {Hinojosa~Jr}}, \ and\ \bibinfo {author} {\bibfnamefont {J.~F.}\ \bibnamefont {Weaver}},\ }\href@noop {} {\bibfield  {journal} {\bibinfo  {journal} {Surface Science}\ }\textbf {\bibinfo {volume} {602}},\ \bibinfo {pages} {3116} (\bibinfo {year} {2008})}\BibitemShut {NoStop}%
\bibitem [{\citenamefont {Weaver}\ \emph {et~al.}(2005)\citenamefont {Weaver}, \citenamefont {Chen},\ and\ \citenamefont {Gerrard}}]{weaver2005oxidation}%
  \BibitemOpen
  \bibfield  {author} {\bibinfo {author} {\bibfnamefont {J.~F.}\ \bibnamefont {Weaver}}, \bibinfo {author} {\bibfnamefont {J.-J.}\ \bibnamefont {Chen}}, \ and\ \bibinfo {author} {\bibfnamefont {A.~L.}\ \bibnamefont {Gerrard}},\ }\href@noop {} {\bibfield  {journal} {\bibinfo  {journal} {Surface Science}\ }\textbf {\bibinfo {volume} {592}},\ \bibinfo {pages} {83} (\bibinfo {year} {2005})}\BibitemShut {NoStop}%
\bibitem [{\citenamefont {van Spronsen}\ \emph {et~al.}(2017)\citenamefont {van Spronsen}, \citenamefont {Frenken},\ and\ \citenamefont {Groot}}]{vanSpronsen2017NC}%
  \BibitemOpen
  \bibfield  {author} {\bibinfo {author} {\bibfnamefont {M.~A.}\ \bibnamefont {van Spronsen}}, \bibinfo {author} {\bibfnamefont {J.~W.~M.}\ \bibnamefont {Frenken}}, \ and\ \bibinfo {author} {\bibfnamefont {I.~M.~N.}\ \bibnamefont {Groot}},\ }\href {\doibase 10.1038/s41467-017-00643-z} {\bibfield  {journal} {\bibinfo  {journal} {Nature Communications}\ }\textbf {\bibinfo {volume} {8}},\ \bibinfo {pages} {429} (\bibinfo {year} {2017})}\BibitemShut {NoStop}%
\bibitem [{\citenamefont {Cai}\ \emph {et~al.}(2024)\citenamefont {Cai}, \citenamefont {Wei}, \citenamefont {Liu}, \citenamefont {Xue}, \citenamefont {Chen}, \citenamefont {Hu}, \citenamefont {Zang}, \citenamefont {Wang}, \citenamefont {Shi}, \citenamefont {Qin} \emph {et~al.}}]{wang2024NM}%
  \BibitemOpen
  \bibfield  {author} {\bibinfo {author} {\bibfnamefont {J.}~\bibnamefont {Cai}}, \bibinfo {author} {\bibfnamefont {L.}~\bibnamefont {Wei}}, \bibinfo {author} {\bibfnamefont {J.}~\bibnamefont {Liu}}, \bibinfo {author} {\bibfnamefont {C.}~\bibnamefont {Xue}}, \bibinfo {author} {\bibfnamefont {Z.}~\bibnamefont {Chen}}, \bibinfo {author} {\bibfnamefont {Y.}~\bibnamefont {Hu}}, \bibinfo {author} {\bibfnamefont {Y.}~\bibnamefont {Zang}}, \bibinfo {author} {\bibfnamefont {M.}~\bibnamefont {Wang}}, \bibinfo {author} {\bibfnamefont {W.}~\bibnamefont {Shi}}, \bibinfo {author} {\bibfnamefont {T.}~\bibnamefont {Qin}},  \emph {et~al.},\ }\href@noop {} {\bibfield  {journal} {\bibinfo  {journal} {Nature materials}\ }\textbf {\bibinfo {volume} {23}},\ \bibinfo {pages} {1654} (\bibinfo {year} {2024})}\BibitemShut {NoStop}%
\bibitem [{\citenamefont {Chaston}(1965)}]{chaston1965reactions}%
  \BibitemOpen
  \bibfield  {author} {\bibinfo {author} {\bibfnamefont {J.}~\bibnamefont {Chaston}},\ }\href@noop {} {\bibfield  {journal} {\bibinfo  {journal} {Platinum Metals Review}\ }\textbf {\bibinfo {volume} {9}},\ \bibinfo {pages} {51} (\bibinfo {year} {1965})}\BibitemShut {NoStop}%
\bibitem [{\citenamefont {Maxwell}(1864)}]{maxwell1864xlv}%
  \BibitemOpen
  \bibfield  {author} {\bibinfo {author} {\bibfnamefont {J.~C.}\ \bibnamefont {Maxwell}},\ }\href@noop {} {\bibfield  {journal} {\bibinfo  {journal} {The London, Edinburgh, and Dublin Philosophical Magazine and Journal of Science}\ }\textbf {\bibinfo {volume} {27}},\ \bibinfo {pages} {250} (\bibinfo {year} {1864})}\BibitemShut {NoStop}%
\bibitem [{\citenamefont {Lubensky}\ \emph {et~al.}(2015)\citenamefont {Lubensky}, \citenamefont {Kane}, \citenamefont {Mao}, \citenamefont {Souslov},\ and\ \citenamefont {Sun}}]{Lubensky2015RPP}%
  \BibitemOpen
  \bibfield  {author} {\bibinfo {author} {\bibfnamefont {T.}~\bibnamefont {Lubensky}}, \bibinfo {author} {\bibfnamefont {C.}~\bibnamefont {Kane}}, \bibinfo {author} {\bibfnamefont {X.}~\bibnamefont {Mao}}, \bibinfo {author} {\bibfnamefont {A.}~\bibnamefont {Souslov}}, \ and\ \bibinfo {author} {\bibfnamefont {K.}~\bibnamefont {Sun}},\ }\href@noop {} {\bibfield  {journal} {\bibinfo  {journal} {Reports on Progress in Physics}\ }\textbf {\bibinfo {volume} {78}},\ \bibinfo {pages} {073901} (\bibinfo {year} {2015})}\BibitemShut {NoStop}%
\bibitem [{\citenamefont {Tan}\ and\ \citenamefont {Souslov}(2025)}]{tan2025classifying}%
  \BibitemOpen
  \bibfield  {author} {\bibinfo {author} {\bibfnamefont {I.}~\bibnamefont {Tan}}\ and\ \bibinfo {author} {\bibfnamefont {A.}~\bibnamefont {Souslov}},\ }\href@noop {} {\bibfield  {journal} {\bibinfo  {journal} {New Journal of Physics}\ }\textbf {\bibinfo {volume} {27}},\ \bibinfo {pages} {055002} (\bibinfo {year} {2025})}\BibitemShut {NoStop}%
\bibitem [{\citenamefont {Zhang}\ and\ \citenamefont {Mao}(2018)}]{zhang2018fracturing}%
  \BibitemOpen
  \bibfield  {author} {\bibinfo {author} {\bibfnamefont {L.}~\bibnamefont {Zhang}}\ and\ \bibinfo {author} {\bibfnamefont {X.}~\bibnamefont {Mao}},\ }\href@noop {} {\bibfield  {journal} {\bibinfo  {journal} {New Journal of Physics}\ }\textbf {\bibinfo {volume} {20}},\ \bibinfo {pages} {063034} (\bibinfo {year} {2018})}\BibitemShut {NoStop}%
\bibitem [{\citenamefont {He}\ \emph {et~al.}(2021)\citenamefont {He}, \citenamefont {Zhou}, \citenamefont {Ye}, \citenamefont {Cho}, \citenamefont {Jeong}, \citenamefont {Meng},\ and\ \citenamefont {Wang}}]{he2021moire}%
  \BibitemOpen
  \bibfield  {author} {\bibinfo {author} {\bibfnamefont {F.}~\bibnamefont {He}}, \bibinfo {author} {\bibfnamefont {Y.}~\bibnamefont {Zhou}}, \bibinfo {author} {\bibfnamefont {Z.}~\bibnamefont {Ye}}, \bibinfo {author} {\bibfnamefont {S.-H.}\ \bibnamefont {Cho}}, \bibinfo {author} {\bibfnamefont {J.}~\bibnamefont {Jeong}}, \bibinfo {author} {\bibfnamefont {X.}~\bibnamefont {Meng}}, \ and\ \bibinfo {author} {\bibfnamefont {Y.}~\bibnamefont {Wang}},\ }\href@noop {} {\bibfield  {journal} {\bibinfo  {journal} {ACS nano}\ }\textbf {\bibinfo {volume} {15}},\ \bibinfo {pages} {5944} (\bibinfo {year} {2021})}\BibitemShut {NoStop}%
\bibitem [{\citenamefont {Wyart}(2005)}]{Wyart2005}%
  \BibitemOpen
  \bibfield  {author} {\bibinfo {author} {\bibfnamefont {M.}~\bibnamefont {Wyart}},\ }\href {\doibase 10.1051/anphys:2006003} {\bibfield  {journal} {\bibinfo  {journal} {Annales de Physique}\ }\textbf {\bibinfo {volume} {30}},\ \bibinfo {pages} {1} (\bibinfo {year} {2005})}\BibitemShut {NoStop}%
\bibitem [{\citenamefont {Zaccone}\ and\ \citenamefont {Scossa-Romano}(2011)}]{Zaccone5}%
  \BibitemOpen
  \bibfield  {author} {\bibinfo {author} {\bibfnamefont {A.}~\bibnamefont {Zaccone}}\ and\ \bibinfo {author} {\bibfnamefont {E.}~\bibnamefont {Scossa-Romano}},\ }\href {\doibase 10.1103/PhysRevB.83.184205} {\bibfield  {journal} {\bibinfo  {journal} {Phys. Rev. B}\ }\textbf {\bibinfo {volume} {83}},\ \bibinfo {pages} {184205} (\bibinfo {year} {2011})}\BibitemShut {NoStop}%
\bibitem [{\citenamefont {Zhang}\ \emph {et~al.}(2009)\citenamefont {Zhang}, \citenamefont {Xu}, \citenamefont {Chen}, \citenamefont {Yunker}, \citenamefont {Alsayed}, \citenamefont {Aptowicz}, \citenamefont {Habdas}, \citenamefont {Liu}, \citenamefont {Nagel},\ and\ \citenamefont {Yodh}}]{Zhang2009N}%
  \BibitemOpen
  \bibfield  {author} {\bibinfo {author} {\bibfnamefont {Z.}~\bibnamefont {Zhang}}, \bibinfo {author} {\bibfnamefont {N.}~\bibnamefont {Xu}}, \bibinfo {author} {\bibfnamefont {D.~T.~N.}\ \bibnamefont {Chen}}, \bibinfo {author} {\bibfnamefont {P.}~\bibnamefont {Yunker}}, \bibinfo {author} {\bibfnamefont {A.~M.}\ \bibnamefont {Alsayed}}, \bibinfo {author} {\bibfnamefont {K.~B.}\ \bibnamefont {Aptowicz}}, \bibinfo {author} {\bibfnamefont {P.}~\bibnamefont {Habdas}}, \bibinfo {author} {\bibfnamefont {A.~J.}\ \bibnamefont {Liu}}, \bibinfo {author} {\bibfnamefont {S.~R.}\ \bibnamefont {Nagel}}, \ and\ \bibinfo {author} {\bibfnamefont {A.~G.}\ \bibnamefont {Yodh}},\ }\href {\doibase 10.1038/nature07998} {\bibfield  {journal} {\bibinfo  {journal} {Nature}\ }\textbf {\bibinfo {volume} {459}},\ \bibinfo {pages} {230} (\bibinfo {year} {2009})}\BibitemShut {NoStop}%
\bibitem [{\citenamefont {Fruchart}\ \emph {et~al.}(2020)\citenamefont {Fruchart}, \citenamefont {Zhou},\ and\ \citenamefont {Vitelli}}]{Fruchart2020N}%
  \BibitemOpen
  \bibfield  {author} {\bibinfo {author} {\bibfnamefont {M.}~\bibnamefont {Fruchart}}, \bibinfo {author} {\bibfnamefont {Y.}~\bibnamefont {Zhou}}, \ and\ \bibinfo {author} {\bibfnamefont {V.}~\bibnamefont {Vitelli}},\ }\href@noop {} {\bibfield  {journal} {\bibinfo  {journal} {Nature}\ }\textbf {\bibinfo {volume} {577}},\ \bibinfo {pages} {636} (\bibinfo {year} {2020})}\BibitemShut {NoStop}%
\bibitem [{\citenamefont {Lei}\ \emph {et~al.}(2021)\citenamefont {Lei}, \citenamefont {Zheng}, \citenamefont {Tang}, \citenamefont {Wan}, \citenamefont {Ni},\ and\ \citenamefont {Ma}}]{Lei2021PRL}%
  \BibitemOpen
  \bibfield  {author} {\bibinfo {author} {\bibfnamefont {Q.-L.}\ \bibnamefont {Lei}}, \bibinfo {author} {\bibfnamefont {W.}~\bibnamefont {Zheng}}, \bibinfo {author} {\bibfnamefont {F.}~\bibnamefont {Tang}}, \bibinfo {author} {\bibfnamefont {X.}~\bibnamefont {Wan}}, \bibinfo {author} {\bibfnamefont {R.}~\bibnamefont {Ni}}, \ and\ \bibinfo {author} {\bibfnamefont {Y.-q.}\ \bibnamefont {Ma}},\ }\href {\doibase 10.1103/PhysRevLett.127.018001} {\bibfield  {journal} {\bibinfo  {journal} {Phys. Rev. Lett.}\ }\textbf {\bibinfo {volume} {127}},\ \bibinfo {pages} {018001} (\bibinfo {year} {2021})}\BibitemShut {NoStop}%
\bibitem [{\citenamefont {Cui}\ \emph {et~al.}(2019)\citenamefont {Cui}, \citenamefont {Zaccone},\ and\ \citenamefont {Rodney}}]{Zaccone2}%
  \BibitemOpen
  \bibfield  {author} {\bibinfo {author} {\bibfnamefont {B.}~\bibnamefont {Cui}}, \bibinfo {author} {\bibfnamefont {A.}~\bibnamefont {Zaccone}}, \ and\ \bibinfo {author} {\bibfnamefont {D.}~\bibnamefont {Rodney}},\ }\href {\doibase 10.1063/1.5129025} {\bibfield  {journal} {\bibinfo  {journal} {The Journal of Chemical Physics}\ }\textbf {\bibinfo {volume} {151}},\ \bibinfo {pages} {224509} (\bibinfo {year} {2019})}\BibitemShut {NoStop}%
\bibitem [{\citenamefont {Kane}\ and\ \citenamefont {Lubensky}(2014)}]{kane2014NP}%
  \BibitemOpen
  \bibfield  {author} {\bibinfo {author} {\bibfnamefont {C.}~\bibnamefont {Kane}}\ and\ \bibinfo {author} {\bibfnamefont {T.}~\bibnamefont {Lubensky}},\ }\href@noop {} {\bibfield  {journal} {\bibinfo  {journal} {Nature Physics}\ }\textbf {\bibinfo {volume} {10}},\ \bibinfo {pages} {39} (\bibinfo {year} {2014})}\BibitemShut {NoStop}%
\bibitem [{\citenamefont {Zhou}\ \emph {et~al.}(2019)\citenamefont {Zhou}, \citenamefont {Zhang},\ and\ \citenamefont {Mao}}]{Zhou2019PRX}%
  \BibitemOpen
  \bibfield  {author} {\bibinfo {author} {\bibfnamefont {D.}~\bibnamefont {Zhou}}, \bibinfo {author} {\bibfnamefont {L.}~\bibnamefont {Zhang}}, \ and\ \bibinfo {author} {\bibfnamefont {X.}~\bibnamefont {Mao}},\ }\href {\doibase 10.1103/PhysRevX.9.021054} {\bibfield  {journal} {\bibinfo  {journal} {Phys. Rev. X}\ }\textbf {\bibinfo {volume} {9}},\ \bibinfo {pages} {021054} (\bibinfo {year} {2019})}\BibitemShut {NoStop}%
\bibitem [{\citenamefont {Bertoldi}\ \emph {et~al.}(2017)\citenamefont {Bertoldi}, \citenamefont {Vitelli}, \citenamefont {Christensen},\ and\ \citenamefont {van Hecke}}]{Bertoldi2017}%
  \BibitemOpen
  \bibfield  {author} {\bibinfo {author} {\bibfnamefont {K.}~\bibnamefont {Bertoldi}}, \bibinfo {author} {\bibfnamefont {V.}~\bibnamefont {Vitelli}}, \bibinfo {author} {\bibfnamefont {J.}~\bibnamefont {Christensen}}, \ and\ \bibinfo {author} {\bibfnamefont {M.}~\bibnamefont {van Hecke}},\ }\href {\doibase 10.1038/natrevmats.2017.66} {\bibfield  {journal} {\bibinfo  {journal} {Nature Reviews Materials}\ }\textbf {\bibinfo {volume} {2}},\ \bibinfo {pages} {17066} (\bibinfo {year} {2017})}\BibitemShut {NoStop}%
\bibitem [{\citenamefont {Liu}\ \emph {et~al.}(2022)\citenamefont {Liu}, \citenamefont {B{\o}jesen}, \citenamefont {Tabor}, \citenamefont {Mudie}, \citenamefont {Zaccone}, \citenamefont {Harrowell},\ and\ \citenamefont {Petersen}}]{Zaccone3}%
  \BibitemOpen
  \bibfield  {author} {\bibinfo {author} {\bibfnamefont {A.~C.}\ \bibnamefont {Liu}}, \bibinfo {author} {\bibfnamefont {E.~D.}\ \bibnamefont {B{\o}jesen}}, \bibinfo {author} {\bibfnamefont {R.~F.}\ \bibnamefont {Tabor}}, \bibinfo {author} {\bibfnamefont {S.~T.}\ \bibnamefont {Mudie}}, \bibinfo {author} {\bibfnamefont {A.}~\bibnamefont {Zaccone}}, \bibinfo {author} {\bibfnamefont {P.}~\bibnamefont {Harrowell}}, \ and\ \bibinfo {author} {\bibfnamefont {T.~C.}\ \bibnamefont {Petersen}},\ }\href {\doibase 10.1126/sciadv.abn0681} {\bibfield  {journal} {\bibinfo  {journal} {Science Advances}\ }\textbf {\bibinfo {volume} {8}},\ \bibinfo {pages} {eabn0681} (\bibinfo {year} {2022})}\BibitemShut {NoStop}%
\bibitem [{\citenamefont {McGilly}\ \emph {et~al.}(2020)\citenamefont {McGilly}, \citenamefont {Kerelsky}, \citenamefont {Finney}, \citenamefont {Shapovalov}, \citenamefont {Shih}, \citenamefont {Ghiotto}, \citenamefont {Zeng}, \citenamefont {Moore}, \citenamefont {Wu}, \citenamefont {Bai} \emph {et~al.}}]{mcgilly2020visualization}%
  \BibitemOpen
  \bibfield  {author} {\bibinfo {author} {\bibfnamefont {L.~J.}\ \bibnamefont {McGilly}}, \bibinfo {author} {\bibfnamefont {A.}~\bibnamefont {Kerelsky}}, \bibinfo {author} {\bibfnamefont {N.~R.}\ \bibnamefont {Finney}}, \bibinfo {author} {\bibfnamefont {K.}~\bibnamefont {Shapovalov}}, \bibinfo {author} {\bibfnamefont {E.-M.}\ \bibnamefont {Shih}}, \bibinfo {author} {\bibfnamefont {A.}~\bibnamefont {Ghiotto}}, \bibinfo {author} {\bibfnamefont {Y.}~\bibnamefont {Zeng}}, \bibinfo {author} {\bibfnamefont {S.~L.}\ \bibnamefont {Moore}}, \bibinfo {author} {\bibfnamefont {W.}~\bibnamefont {Wu}}, \bibinfo {author} {\bibfnamefont {Y.}~\bibnamefont {Bai}},  \emph {et~al.},\ }\href@noop {} {\bibfield  {journal} {\bibinfo  {journal} {Nature Nanotechnology}\ }\textbf {\bibinfo {volume} {15}},\ \bibinfo {pages} {580} (\bibinfo {year} {2020})}\BibitemShut {NoStop}%
\bibitem [{\citenamefont {Vashishta}\ \emph {et~al.}(1990)\citenamefont {Vashishta}, \citenamefont {Kalia}, \citenamefont {Rino},\ and\ \citenamefont {Ebbsj\"o}}]{Vashishta1990PRB}%
  \BibitemOpen
  \bibfield  {author} {\bibinfo {author} {\bibfnamefont {P.}~\bibnamefont {Vashishta}}, \bibinfo {author} {\bibfnamefont {R.~K.}\ \bibnamefont {Kalia}}, \bibinfo {author} {\bibfnamefont {J.~P.}\ \bibnamefont {Rino}}, \ and\ \bibinfo {author} {\bibfnamefont {I.}~\bibnamefont {Ebbsj\"o}},\ }\href {\doibase 10.1103/PhysRevB.41.12197} {\bibfield  {journal} {\bibinfo  {journal} {Physical Review B}\ }\textbf {\bibinfo {volume} {41}},\ \bibinfo {pages} {12197} (\bibinfo {year} {1990})}\BibitemShut {NoStop}%
\bibitem [{\citenamefont {Chen}\ \emph {et~al.}(2019)\citenamefont {Chen}, \citenamefont {Li},\ and\ \citenamefont {Yang}}]{Chen2019FOP}%
  \BibitemOpen
  \bibfield  {author} {\bibinfo {author} {\bibfnamefont {Q.}~\bibnamefont {Chen}}, \bibinfo {author} {\bibfnamefont {W.}~\bibnamefont {Li}}, \ and\ \bibinfo {author} {\bibfnamefont {Y.}~\bibnamefont {Yang}},\ }\href {\doibase 10.1007/s11467-019-0900-9} {\bibfield  {journal} {\bibinfo  {journal} {Frontiers of Physics}\ }\textbf {\bibinfo {volume} {14}},\ \bibinfo {pages} {53604} (\bibinfo {year} {2019})}\BibitemShut {NoStop}%
\bibitem [{\citenamefont {Hamilton}\ and\ \citenamefont {Strachan}(2023)}]{PhysRevMaterials.7.075601}%
  \BibitemOpen
  \bibfield  {author} {\bibinfo {author} {\bibfnamefont {B.~W.}\ \bibnamefont {Hamilton}}\ and\ \bibinfo {author} {\bibfnamefont {A.}~\bibnamefont {Strachan}},\ }\href {\doibase 10.1103/PhysRevMaterials.7.075601} {\bibfield  {journal} {\bibinfo  {journal} {Phys. Rev. Mater.}\ }\textbf {\bibinfo {volume} {7}},\ \bibinfo {pages} {075601} (\bibinfo {year} {2023})}\BibitemShut {NoStop}%
\bibitem [{\citenamefont {Paulose}\ \emph {et~al.}(2015)\citenamefont {Paulose}, \citenamefont {Chen},\ and\ \citenamefont {Vitelli}}]{paulose2015topological}%
  \BibitemOpen
  \bibfield  {author} {\bibinfo {author} {\bibfnamefont {J.}~\bibnamefont {Paulose}}, \bibinfo {author} {\bibfnamefont {B.~G.-g.}\ \bibnamefont {Chen}}, \ and\ \bibinfo {author} {\bibfnamefont {V.}~\bibnamefont {Vitelli}},\ }\href@noop {} {\bibfield  {journal} {\bibinfo  {journal} {Nature Physics}\ }\textbf {\bibinfo {volume} {11}},\ \bibinfo {pages} {153} (\bibinfo {year} {2015})}\BibitemShut {NoStop}%
\bibitem [{\citenamefont {Ma}\ \emph {et~al.}(2023)\citenamefont {Ma}, \citenamefont {Tang}, \citenamefont {Shi}, \citenamefont {Wu}, \citenamefont {Yang}, \citenamefont {Zhou}, \citenamefont {Yao},\ and\ \citenamefont {Li}}]{Ma2023PRL}%
  \BibitemOpen
  \bibfield  {author} {\bibinfo {author} {\bibfnamefont {F.}~\bibnamefont {Ma}}, \bibinfo {author} {\bibfnamefont {Z.}~\bibnamefont {Tang}}, \bibinfo {author} {\bibfnamefont {X.}~\bibnamefont {Shi}}, \bibinfo {author} {\bibfnamefont {Y.}~\bibnamefont {Wu}}, \bibinfo {author} {\bibfnamefont {J.}~\bibnamefont {Yang}}, \bibinfo {author} {\bibfnamefont {D.}~\bibnamefont {Zhou}}, \bibinfo {author} {\bibfnamefont {Y.}~\bibnamefont {Yao}}, \ and\ \bibinfo {author} {\bibfnamefont {F.}~\bibnamefont {Li}},\ }\href {\doibase 10.1103/PhysRevLett.131.046101} {\bibfield  {journal} {\bibinfo  {journal} {Phys. Rev. Lett.}\ }\textbf {\bibinfo {volume} {131}},\ \bibinfo {pages} {046101} (\bibinfo {year} {2023})}\BibitemShut {NoStop}%
\bibitem [{\citenamefont {Milkus}\ and\ \citenamefont {Zaccone}(2016)}]{Zaccone4}%
  \BibitemOpen
  \bibfield  {author} {\bibinfo {author} {\bibfnamefont {R.}~\bibnamefont {Milkus}}\ and\ \bibinfo {author} {\bibfnamefont {A.}~\bibnamefont {Zaccone}},\ }\href {\doibase 10.1103/PhysRevB.93.094204} {\bibfield  {journal} {\bibinfo  {journal} {Phys. Rev. B}\ }\textbf {\bibinfo {volume} {93}},\ \bibinfo {pages} {094204} (\bibinfo {year} {2016})}\BibitemShut {NoStop}%
\bibitem [{\citenamefont {Zaccone}(2023)}]{zaccone1}%
  \BibitemOpen
  \bibfield  {author} {\bibinfo {author} {\bibfnamefont {A.}~\bibnamefont {Zaccone}},\ }\href {\doibase 10.1007/978-3-031-24706-4} {\emph {\bibinfo {title} {Theory of Disordered Solids: From Atomistic Dynamics to Mechanical, Vibrational, and Thermal Properties}}},\ \bibinfo {series} {Lecture Notes in Physics}, Vol.\ \bibinfo {volume} {1015}\ (\bibinfo  {publisher} {Springer Nature Switzerland},\ \bibinfo {address} {Cham, Switzerland},\ \bibinfo {year} {2023})\BibitemShut {NoStop}%
\bibitem [{\citenamefont {Chen}\ \emph {et~al.}(2016)\citenamefont {Chen}, \citenamefont {Liu}, \citenamefont {Evans}, \citenamefont {Paulose}, \citenamefont {Cohen}, \citenamefont {Vitelli},\ and\ \citenamefont {Santangelo}}]{PhysRevLett.116.135501}%
  \BibitemOpen
  \bibfield  {author} {\bibinfo {author} {\bibfnamefont {B.~G.-g.}\ \bibnamefont {Chen}}, \bibinfo {author} {\bibfnamefont {B.}~\bibnamefont {Liu}}, \bibinfo {author} {\bibfnamefont {A.~A.}\ \bibnamefont {Evans}}, \bibinfo {author} {\bibfnamefont {J.}~\bibnamefont {Paulose}}, \bibinfo {author} {\bibfnamefont {I.}~\bibnamefont {Cohen}}, \bibinfo {author} {\bibfnamefont {V.}~\bibnamefont {Vitelli}}, \ and\ \bibinfo {author} {\bibfnamefont {C.~D.}\ \bibnamefont {Santangelo}},\ }\href {\doibase 10.1103/PhysRevLett.116.135501} {\bibfield  {journal} {\bibinfo  {journal} {Phys. Rev. Lett.}\ }\textbf {\bibinfo {volume} {116}},\ \bibinfo {pages} {135501} (\bibinfo {year} {2016})}\BibitemShut {NoStop}%
\bibitem [{\citenamefont {Baardink}\ \emph {et~al.}(2018)\citenamefont {Baardink}, \citenamefont {Souslov}, \citenamefont {Paulose},\ and\ \citenamefont {Vitelli}}]{baardink2018localizing}%
  \BibitemOpen
  \bibfield  {author} {\bibinfo {author} {\bibfnamefont {G.}~\bibnamefont {Baardink}}, \bibinfo {author} {\bibfnamefont {A.}~\bibnamefont {Souslov}}, \bibinfo {author} {\bibfnamefont {J.}~\bibnamefont {Paulose}}, \ and\ \bibinfo {author} {\bibfnamefont {V.}~\bibnamefont {Vitelli}},\ }\href@noop {} {\bibfield  {journal} {\bibinfo  {journal} {Proceedings of the National Academy of Sciences}\ }\textbf {\bibinfo {volume} {115}},\ \bibinfo {pages} {489} (\bibinfo {year} {2018})}\BibitemShut {NoStop}%
\bibitem [{\citenamefont {Zhou}\ \emph {et~al.}(2018)\citenamefont {Zhou}, \citenamefont {Zhang},\ and\ \citenamefont {Mao}}]{Zhou2018PRL}%
  \BibitemOpen
  \bibfield  {author} {\bibinfo {author} {\bibfnamefont {D.}~\bibnamefont {Zhou}}, \bibinfo {author} {\bibfnamefont {L.}~\bibnamefont {Zhang}}, \ and\ \bibinfo {author} {\bibfnamefont {X.}~\bibnamefont {Mao}},\ }\href {\doibase 10.1103/PhysRevLett.120.068003} {\bibfield  {journal} {\bibinfo  {journal} {Phys. Rev. Lett.}\ }\textbf {\bibinfo {volume} {120}},\ \bibinfo {pages} {068003} (\bibinfo {year} {2018})}\BibitemShut {NoStop}%
\bibitem [{\citenamefont {Chen}\ \emph {et~al.}(2025)\citenamefont {Chen}, \citenamefont {McInerney}, \citenamefont {Krause}, \citenamefont {Schneider}, \citenamefont {Wegener},\ and\ \citenamefont {Mao}}]{Mao2025PRL}%
  \BibitemOpen
  \bibfield  {author} {\bibinfo {author} {\bibfnamefont {Y.}~\bibnamefont {Chen}}, \bibinfo {author} {\bibfnamefont {J.~P.}\ \bibnamefont {McInerney}}, \bibinfo {author} {\bibfnamefont {P.~N.}\ \bibnamefont {Krause}}, \bibinfo {author} {\bibfnamefont {J.~L.~G.}\ \bibnamefont {Schneider}}, \bibinfo {author} {\bibfnamefont {M.}~\bibnamefont {Wegener}}, \ and\ \bibinfo {author} {\bibfnamefont {X.}~\bibnamefont {Mao}},\ }\href {\doibase 10.1103/PhysRevLett.134.086101} {\bibfield  {journal} {\bibinfo  {journal} {Phys. Rev. Lett.}\ }\textbf {\bibinfo {volume} {134}},\ \bibinfo {pages} {086101} (\bibinfo {year} {2025})}\BibitemShut {NoStop}%
\bibitem [{\citenamefont {Tang}\ \emph {et~al.}(2024)\citenamefont {Tang}, \citenamefont {Ma}, \citenamefont {Li}, \citenamefont {Yao},\ and\ \citenamefont {Zhou}}]{tang2024PRL}%
  \BibitemOpen
  \bibfield  {author} {\bibinfo {author} {\bibfnamefont {Z.}~\bibnamefont {Tang}}, \bibinfo {author} {\bibfnamefont {F.}~\bibnamefont {Ma}}, \bibinfo {author} {\bibfnamefont {F.}~\bibnamefont {Li}}, \bibinfo {author} {\bibfnamefont {Y.}~\bibnamefont {Yao}}, \ and\ \bibinfo {author} {\bibfnamefont {D.}~\bibnamefont {Zhou}},\ }\href {\doibase 10.1103/PhysRevLett.133.106101} {\bibfield  {journal} {\bibinfo  {journal} {Phys. Rev. Lett.}\ }\textbf {\bibinfo {volume} {133}},\ \bibinfo {pages} {106101} (\bibinfo {year} {2024})}\BibitemShut {NoStop}%
\bibitem [{\citenamefont {Stenull}\ and\ \citenamefont {Lubensky}(2019)}]{Lubensky2019PRL}%
  \BibitemOpen
  \bibfield  {author} {\bibinfo {author} {\bibfnamefont {O.}~\bibnamefont {Stenull}}\ and\ \bibinfo {author} {\bibfnamefont {T.~C.}\ \bibnamefont {Lubensky}},\ }\href {\doibase 10.1103/PhysRevLett.122.248002} {\bibfield  {journal} {\bibinfo  {journal} {Phys. Rev. Lett.}\ }\textbf {\bibinfo {volume} {122}},\ \bibinfo {pages} {248002} (\bibinfo {year} {2019})}\BibitemShut {NoStop}%
\bibitem [{\citenamefont {Rumble}(2024)}]{crc_handbook_2024}%
  \BibitemOpen
  \bibinfo {editor} {\bibfnamefont {J.~R.}\ \bibnamefont {Rumble}},\ ed.,\ \href@noop {} {\emph {\bibinfo {title} {CRC Handbook of Chemistry and Physics}}},\ \bibinfo {edition} {105th}\ ed.\ (\bibinfo  {publisher} {CRC Press},\ \bibinfo {address} {Boca Raton, FL},\ \bibinfo {year} {2024})\BibitemShut {NoStop}%
\bibitem [{\citenamefont {Kittel}(2004)}]{kittel2004introduction}%
  \BibitemOpen
  \bibfield  {author} {\bibinfo {author} {\bibfnamefont {C.}~\bibnamefont {Kittel}},\ }\href@noop {} {\emph {\bibinfo {title} {Introduction to Solid State Physics}}},\ \bibinfo {edition} {8th}\ ed.\ (\bibinfo  {publisher} {John Wiley \& Sons},\ \bibinfo {address} {Hoboken, NJ},\ \bibinfo {year} {2004})\BibitemShut {NoStop}%
\bibitem [{\citenamefont {Sch\"onecker}\ \emph {et~al.}(2018)\citenamefont {Sch\"onecker}, \citenamefont {Li}, \citenamefont {Richter},\ and\ \citenamefont {Vitos}}]{PhysRevB.97.224305}%
  \BibitemOpen
  \bibfield  {author} {\bibinfo {author} {\bibfnamefont {S.}~\bibnamefont {Sch\"onecker}}, \bibinfo {author} {\bibfnamefont {X.}~\bibnamefont {Li}}, \bibinfo {author} {\bibfnamefont {M.}~\bibnamefont {Richter}}, \ and\ \bibinfo {author} {\bibfnamefont {L.}~\bibnamefont {Vitos}},\ }\href {\doibase 10.1103/PhysRevB.97.224305} {\bibfield  {journal} {\bibinfo  {journal} {Phys. Rev. B}\ }\textbf {\bibinfo {volume} {97}},\ \bibinfo {pages} {224305} (\bibinfo {year} {2018})}\BibitemShut {NoStop}%
\bibitem [{\citenamefont {Florijn}\ \emph {et~al.}(2014)\citenamefont {Florijn}, \citenamefont {Coulais},\ and\ \citenamefont {van Hecke}}]{Florijn2014}%
  \BibitemOpen
  \bibfield  {author} {\bibinfo {author} {\bibfnamefont {B.}~\bibnamefont {Florijn}}, \bibinfo {author} {\bibfnamefont {C.}~\bibnamefont {Coulais}}, \ and\ \bibinfo {author} {\bibfnamefont {M.}~\bibnamefont {van Hecke}},\ }\href {\doibase 10.1103/PhysRevLett.113.175503} {\bibfield  {journal} {\bibinfo  {journal} {Physical Review Letters}\ }\textbf {\bibinfo {volume} {113}},\ \bibinfo {pages} {175503} (\bibinfo {year} {2014})}\BibitemShut {NoStop}%
\bibitem [{\citenamefont {Coulais}\ \emph {et~al.}(2016)\citenamefont {Coulais}, \citenamefont {Teomy}, \citenamefont {de~Reus}, \citenamefont {Shokef},\ and\ \citenamefont {van Hecke}}]{Coulais2016}%
  \BibitemOpen
  \bibfield  {author} {\bibinfo {author} {\bibfnamefont {C.}~\bibnamefont {Coulais}}, \bibinfo {author} {\bibfnamefont {E.}~\bibnamefont {Teomy}}, \bibinfo {author} {\bibfnamefont {K.}~\bibnamefont {de~Reus}}, \bibinfo {author} {\bibfnamefont {Y.}~\bibnamefont {Shokef}}, \ and\ \bibinfo {author} {\bibfnamefont {M.}~\bibnamefont {van Hecke}},\ }\href {\doibase 10.1038/nature18960} {\bibfield  {journal} {\bibinfo  {journal} {Nature}\ }\textbf {\bibinfo {volume} {535}},\ \bibinfo {pages} {529} (\bibinfo {year} {2016})}\BibitemShut {NoStop}%
\bibitem [{\citenamefont {Yao}\ \emph {et~al.}(2018)\citenamefont {Yao}, \citenamefont {Wang}, \citenamefont {Bao}, \citenamefont {Zhang}, \citenamefont {Zhang}, \citenamefont {Bao}, \citenamefont {Chan}, \citenamefont {Chen}, \citenamefont {Avila}, \citenamefont {Asensio}, \citenamefont {Zhu},\ and\ \citenamefont {Zhou}}]{Yao2018PNAS}%
  \BibitemOpen
  \bibfield  {author} {\bibinfo {author} {\bibfnamefont {W.}~\bibnamefont {Yao}}, \bibinfo {author} {\bibfnamefont {E.}~\bibnamefont {Wang}}, \bibinfo {author} {\bibfnamefont {C.}~\bibnamefont {Bao}}, \bibinfo {author} {\bibfnamefont {Y.}~\bibnamefont {Zhang}}, \bibinfo {author} {\bibfnamefont {K.}~\bibnamefont {Zhang}}, \bibinfo {author} {\bibfnamefont {K.}~\bibnamefont {Bao}}, \bibinfo {author} {\bibfnamefont {C.~K.}\ \bibnamefont {Chan}}, \bibinfo {author} {\bibfnamefont {C.}~\bibnamefont {Chen}}, \bibinfo {author} {\bibfnamefont {J.}~\bibnamefont {Avila}}, \bibinfo {author} {\bibfnamefont {M.~C.}\ \bibnamefont {Asensio}}, \bibinfo {author} {\bibfnamefont {J.}~\bibnamefont {Zhu}}, \ and\ \bibinfo {author} {\bibfnamefont {S.}~\bibnamefont {Zhou}},\ }\href {\doibase 10.1073/pnas.1720865115} {\bibfield  {journal} {\bibinfo  {journal} {Proceedings of the National Academy of Sciences}\ }\textbf {\bibinfo {volume} {115}},\ \bibinfo {pages} {6928} (\bibinfo {year} {2018})}\BibitemShut {NoStop}%
\bibitem [{SIP()}]{SIPtOx}%
  \BibitemOpen
  \href@noop {} {\bibinfo  {journal} {See Supplementary Information for the construction of stretching energy, bending energy, the dynamical matrix for elastic networks, the rescaling of the Langevin equation, and the numerical verification that the system reaches thermal equilibrium via the equipartition theorem}\ }\BibitemShut {NoStop}%
\bibitem [{\citenamefont {LAX}(1966)}]{RevModPhys.38.541}%
  \BibitemOpen
\bibfield  {journal} {  }\bibfield  {author} {\bibinfo {author} {\bibfnamefont {M.}~\bibnamefont {LAX}},\ }\href {\doibase 10.1103/RevModPhys.38.541} {\bibfield  {journal} {\bibinfo  {journal} {Rev. Mod. Phys.}\ }\textbf {\bibinfo {volume} {38}},\ \bibinfo {pages} {541} (\bibinfo {year} {1966})}\BibitemShut {NoStop}%
\bibitem [{\citenamefont {Hohenberg}\ and\ \citenamefont {Halperin}(1977)}]{RevModPhys.49.435}%
  \BibitemOpen
  \bibfield  {author} {\bibinfo {author} {\bibfnamefont {P.~C.}\ \bibnamefont {Hohenberg}}\ and\ \bibinfo {author} {\bibfnamefont {B.~I.}\ \bibnamefont {Halperin}},\ }\href {\doibase 10.1103/RevModPhys.49.435} {\bibfield  {journal} {\bibinfo  {journal} {Rev. Mod. Phys.}\ }\textbf {\bibinfo {volume} {49}},\ \bibinfo {pages} {435} (\bibinfo {year} {1977})}\BibitemShut {NoStop}%
\bibitem [{\citenamefont {Chapman}\ and\ \citenamefont {Cowling}(1970)}]{Chapman1970}%
  \BibitemOpen
  \bibfield  {author} {\bibinfo {author} {\bibfnamefont {S.}~\bibnamefont {Chapman}}\ and\ \bibinfo {author} {\bibfnamefont {T.~G.}\ \bibnamefont {Cowling}},\ }\href {https://www.cambridge.org} {}\bibinfo {edition} {3rd}\ ed.\ (\bibinfo  {publisher} {Cambridge University Press},\ \bibinfo {address} {Cambridge, UK},\ \bibinfo {year} {1970})\BibitemShut {NoStop}%
\bibitem [{\citenamefont {Batchelor}(2000)}]{batchelor2000introduction}%
  \BibitemOpen
  \bibfield  {author} {\bibinfo {author} {\bibfnamefont {G.~K.}\ \bibnamefont {Batchelor}},\ }\href@noop {} {\emph {\bibinfo {title} {An introduction to fluid dynamics}}}\ (\bibinfo  {publisher} {Cambridge university press},\ \bibinfo {year} {2000})\BibitemShut {NoStop}%
\bibitem [{\citenamefont {Kim}\ \emph {et~al.}(2006)\citenamefont {Kim}, \citenamefont {Sumi}, \citenamefont {Takahashi}, \citenamefont {Yokoyama}, \citenamefont {Ito}, \citenamefont {Watanabe}, \citenamefont {Akiyama}, \citenamefont {Kaneko}, \citenamefont {Saito},\ and\ \citenamefont {Funakubo}}]{kim2006metalorganic}%
  \BibitemOpen
  \bibfield  {author} {\bibinfo {author} {\bibfnamefont {Y.~K.}\ \bibnamefont {Kim}}, \bibinfo {author} {\bibfnamefont {A.}~\bibnamefont {Sumi}}, \bibinfo {author} {\bibfnamefont {K.}~\bibnamefont {Takahashi}}, \bibinfo {author} {\bibfnamefont {S.}~\bibnamefont {Yokoyama}}, \bibinfo {author} {\bibfnamefont {S.}~\bibnamefont {Ito}}, \bibinfo {author} {\bibfnamefont {T.}~\bibnamefont {Watanabe}}, \bibinfo {author} {\bibfnamefont {K.}~\bibnamefont {Akiyama}}, \bibinfo {author} {\bibfnamefont {S.}~\bibnamefont {Kaneko}}, \bibinfo {author} {\bibfnamefont {K.}~\bibnamefont {Saito}}, \ and\ \bibinfo {author} {\bibfnamefont {H.}~\bibnamefont {Funakubo}},\ }\href@noop {} {\bibfield  {journal} {\bibinfo  {journal} {Japanese Journal of Applied Physics}\ }\textbf {\bibinfo {volume} {45}},\ \bibinfo {pages} {L36} (\bibinfo {year} {2006})}\BibitemShut {NoStop}%
\bibitem [{\citenamefont {Longo}\ \emph {et~al.}(1971)\citenamefont {Longo}, \citenamefont {Kafalas},\ and\ \citenamefont {Arnott}}]{longo1971structure}%
  \BibitemOpen
  \bibfield  {author} {\bibinfo {author} {\bibfnamefont {J.}~\bibnamefont {Longo}}, \bibinfo {author} {\bibfnamefont {J.}~\bibnamefont {Kafalas}}, \ and\ \bibinfo {author} {\bibfnamefont {R.}~\bibnamefont {Arnott}},\ }\href@noop {} {\bibfield  {journal} {\bibinfo  {journal} {Journal of Solid State Chemistry}\ }\textbf {\bibinfo {volume} {3}},\ \bibinfo {pages} {174} (\bibinfo {year} {1971})}\BibitemShut {NoStop}%
\bibitem [{\citenamefont {Moon}\ \emph {et~al.}(2008)\citenamefont {Moon}, \citenamefont {Jin}, \citenamefont {Kim}, \citenamefont {Choi}, \citenamefont {Lee}, \citenamefont {Yu}, \citenamefont {Cao}, \citenamefont {Sumi}, \citenamefont {Funakubo}, \citenamefont {Bernhard} \emph {et~al.}}]{moon2008dimensionality}%
  \BibitemOpen
  \bibfield  {author} {\bibinfo {author} {\bibfnamefont {S.}~\bibnamefont {Moon}}, \bibinfo {author} {\bibfnamefont {H.}~\bibnamefont {Jin}}, \bibinfo {author} {\bibfnamefont {K.~W.}\ \bibnamefont {Kim}}, \bibinfo {author} {\bibfnamefont {W.}~\bibnamefont {Choi}}, \bibinfo {author} {\bibfnamefont {Y.}~\bibnamefont {Lee}}, \bibinfo {author} {\bibfnamefont {J.}~\bibnamefont {Yu}}, \bibinfo {author} {\bibfnamefont {G.}~\bibnamefont {Cao}}, \bibinfo {author} {\bibfnamefont {A.}~\bibnamefont {Sumi}}, \bibinfo {author} {\bibfnamefont {.~f.~H.}\ \bibnamefont {Funakubo}}, \bibinfo {author} {\bibfnamefont {C.}~\bibnamefont {Bernhard}},  \emph {et~al.},\ }\href@noop {} {\bibfield  {journal} {\bibinfo  {journal} {Physical Review Letters}\ }\textbf {\bibinfo {volume} {101}},\ \bibinfo {pages} {226402} (\bibinfo {year} {2008})}\BibitemShut {NoStop}%
\bibitem [{\citenamefont {Okamoto}\ and\ \citenamefont {Xiao}(2018)}]{okamoto2018transition}%
  \BibitemOpen
  \bibfield  {author} {\bibinfo {author} {\bibfnamefont {S.}~\bibnamefont {Okamoto}}\ and\ \bibinfo {author} {\bibfnamefont {D.}~\bibnamefont {Xiao}},\ }\href@noop {} {\bibfield  {journal} {\bibinfo  {journal} {Journal of the Physical Society of Japan}\ }\textbf {\bibinfo {volume} {87}},\ \bibinfo {pages} {041006} (\bibinfo {year} {2018})}\BibitemShut {NoStop}%
\bibitem [{\citenamefont {Doennig}\ \emph {et~al.}(2013)\citenamefont {Doennig}, \citenamefont {Pickett},\ and\ \citenamefont {Pentcheva}}]{doennig2013massive}%
  \BibitemOpen
  \bibfield  {author} {\bibinfo {author} {\bibfnamefont {D.}~\bibnamefont {Doennig}}, \bibinfo {author} {\bibfnamefont {W.~E.}\ \bibnamefont {Pickett}}, \ and\ \bibinfo {author} {\bibfnamefont {R.}~\bibnamefont {Pentcheva}},\ }\href@noop {} {\bibfield  {journal} {\bibinfo  {journal} {Physical Review Letters}\ }\textbf {\bibinfo {volume} {111}},\ \bibinfo {pages} {126804} (\bibinfo {year} {2013})}\BibitemShut {NoStop}%
\bibitem [{\citenamefont {Rawl}\ \emph {et~al.}(2017)\citenamefont {Rawl}, \citenamefont {Lee}, \citenamefont {Choi}, \citenamefont {Li}, \citenamefont {Chen}, \citenamefont {Baumbach}, \citenamefont {Dela~Cruz}, \citenamefont {Ma},\ and\ \citenamefont {Zhou}}]{rawl2017magnetic}%
  \BibitemOpen
  \bibfield  {author} {\bibinfo {author} {\bibfnamefont {R.}~\bibnamefont {Rawl}}, \bibinfo {author} {\bibfnamefont {M.}~\bibnamefont {Lee}}, \bibinfo {author} {\bibfnamefont {E.~S.}\ \bibnamefont {Choi}}, \bibinfo {author} {\bibfnamefont {G.}~\bibnamefont {Li}}, \bibinfo {author} {\bibfnamefont {K.-W.}\ \bibnamefont {Chen}}, \bibinfo {author} {\bibfnamefont {R.}~\bibnamefont {Baumbach}}, \bibinfo {author} {\bibfnamefont {C.}~\bibnamefont {Dela~Cruz}}, \bibinfo {author} {\bibfnamefont {J.}~\bibnamefont {Ma}}, \ and\ \bibinfo {author} {\bibfnamefont {H.}~\bibnamefont {Zhou}},\ }\href@noop {} {\bibfield  {journal} {\bibinfo  {journal} {Physical Review B}\ }\textbf {\bibinfo {volume} {95}},\ \bibinfo {pages} {174438} (\bibinfo {year} {2017})}\BibitemShut {NoStop}%
\bibitem [{\citenamefont {Sumi}\ \emph {et~al.}(2005)\citenamefont {Sumi}, \citenamefont {Kim}, \citenamefont {Oshima}, \citenamefont {Akiyama}, \citenamefont {Saito},\ and\ \citenamefont {Funakubo}}]{sumi2005mocvd}%
  \BibitemOpen
  \bibfield  {author} {\bibinfo {author} {\bibfnamefont {A.}~\bibnamefont {Sumi}}, \bibinfo {author} {\bibfnamefont {Y.}~\bibnamefont {Kim}}, \bibinfo {author} {\bibfnamefont {N.}~\bibnamefont {Oshima}}, \bibinfo {author} {\bibfnamefont {K.}~\bibnamefont {Akiyama}}, \bibinfo {author} {\bibfnamefont {K.}~\bibnamefont {Saito}}, \ and\ \bibinfo {author} {\bibfnamefont {H.}~\bibnamefont {Funakubo}},\ }\href@noop {} {\bibfield  {journal} {\bibinfo  {journal} {Thin Solid Films}\ }\textbf {\bibinfo {volume} {486}},\ \bibinfo {pages} {182} (\bibinfo {year} {2005})}\BibitemShut {NoStop}%
\bibitem [{\citenamefont {Ma}\ \emph {et~al.}(2014)\citenamefont {Ma}, \citenamefont {Yang}, \citenamefont {Xiao}, \citenamefont {Yang},\ and\ \citenamefont {Sheng}}]{Ma2014}%
  \BibitemOpen
  \bibfield  {author} {\bibinfo {author} {\bibfnamefont {G.}~\bibnamefont {Ma}}, \bibinfo {author} {\bibfnamefont {M.}~\bibnamefont {Yang}}, \bibinfo {author} {\bibfnamefont {S.}~\bibnamefont {Xiao}}, \bibinfo {author} {\bibfnamefont {Z.}~\bibnamefont {Yang}}, \ and\ \bibinfo {author} {\bibfnamefont {P.}~\bibnamefont {Sheng}},\ }\href {\doibase 10.1038/nmat3994} {\bibfield  {journal} {\bibinfo  {journal} {Nature Materials}\ }\textbf {\bibinfo {volume} {13}},\ \bibinfo {pages} {873} (\bibinfo {year} {2014})}\BibitemShut {NoStop}%
\bibitem [{\citenamefont {Li}\ and\ \citenamefont {Assouar}(2016)}]{Li2016}%
  \BibitemOpen
  \bibfield  {author} {\bibinfo {author} {\bibfnamefont {Y.}~\bibnamefont {Li}}\ and\ \bibinfo {author} {\bibfnamefont {B.~M.}\ \bibnamefont {Assouar}},\ }\href {\doibase 10.1063/1.4941338} {\bibfield  {journal} {\bibinfo  {journal} {Applied Physics Letters}\ }\textbf {\bibinfo {volume} {108}},\ \bibinfo {pages} {063502} (\bibinfo {year} {2016})}\BibitemShut {NoStop}%
\bibitem [{\citenamefont {Assouar}\ \emph {et~al.}(2018)\citenamefont {Assouar}, \citenamefont {Liang}, \citenamefont {Wu}, \citenamefont {Cheng}, \citenamefont {Jing},\ and\ \citenamefont {Li}}]{Assouar2018}%
  \BibitemOpen
  \bibfield  {author} {\bibinfo {author} {\bibfnamefont {B.}~\bibnamefont {Assouar}}, \bibinfo {author} {\bibfnamefont {B.}~\bibnamefont {Liang}}, \bibinfo {author} {\bibfnamefont {Y.}~\bibnamefont {Wu}}, \bibinfo {author} {\bibfnamefont {J.-C.}\ \bibnamefont {Cheng}}, \bibinfo {author} {\bibfnamefont {Y.}~\bibnamefont {Jing}}, \ and\ \bibinfo {author} {\bibfnamefont {J.}~\bibnamefont {Li}},\ }\href {\doibase 10.1038/s41578-018-0061-4} {\bibfield  {journal} {\bibinfo  {journal} {Nature Reviews Materials}\ }\textbf {\bibinfo {volume} {3}},\ \bibinfo {pages} {460} (\bibinfo {year} {2018})}\BibitemShut {NoStop}%
\bibitem [{\citenamefont {Oudich}\ \emph {et~al.}(2010)\citenamefont {Oudich}, \citenamefont {Li}, \citenamefont {Assouar},\ and\ \citenamefont {Hou}}]{Oudich2010}%
  \BibitemOpen
  \bibfield  {author} {\bibinfo {author} {\bibfnamefont {M.}~\bibnamefont {Oudich}}, \bibinfo {author} {\bibfnamefont {Y.}~\bibnamefont {Li}}, \bibinfo {author} {\bibfnamefont {B.~M.}\ \bibnamefont {Assouar}}, \ and\ \bibinfo {author} {\bibfnamefont {Z.}~\bibnamefont {Hou}},\ }\href {\doibase 10.1088/1367-2630/12/8/083049} {\bibfield  {journal} {\bibinfo  {journal} {New Journal of Physics}\ }\textbf {\bibinfo {volume} {12}},\ \bibinfo {pages} {083049} (\bibinfo {year} {2010})}\BibitemShut {NoStop}%
\bibitem [{\citenamefont {{Zhou}}\ \emph {et~al.}(2022)\citenamefont {{Zhou}}, \citenamefont {{Zeb Rocklin}}, \citenamefont {{Leamy}},\ and\ \citenamefont {{Yao}}}]{Zhou2022NC}%
  \BibitemOpen
  \bibfield  {author} {\bibinfo {author} {\bibfnamefont {D.}~\bibnamefont {{Zhou}}}, \bibinfo {author} {\bibfnamefont {D.}~\bibnamefont {{Zeb Rocklin}}}, \bibinfo {author} {\bibfnamefont {M.}~\bibnamefont {{Leamy}}}, \ and\ \bibinfo {author} {\bibfnamefont {Y.}~\bibnamefont {{Yao}}},\ }\href {\doibase 10.1038/s41467-022-31084-y} {\bibfield  {journal} {\bibinfo  {journal} {Nature Communications}\ }\textbf {\bibinfo {volume} {13}},\ \bibinfo {pages} {3379} (\bibinfo {year} {2022})}\BibitemShut {NoStop}%
\bibitem [{\citenamefont {Xu}\ \emph {et~al.}(2023)\citenamefont {Xu}, \citenamefont {Shen}, \citenamefont {Jiang}, \citenamefont {Yang}, \citenamefont {He}, \citenamefont {Wu}, \citenamefont {Li},\ and\ \citenamefont {Guo}}]{Xu2023}%
  \BibitemOpen
  \bibfield  {author} {\bibinfo {author} {\bibfnamefont {Z.-Y.}\ \bibnamefont {Xu}}, \bibinfo {author} {\bibfnamefont {Y.-Z.}\ \bibnamefont {Shen}}, \bibinfo {author} {\bibfnamefont {J.-T.}\ \bibnamefont {Jiang}}, \bibinfo {author} {\bibfnamefont {B.-H.}\ \bibnamefont {Yang}}, \bibinfo {author} {\bibfnamefont {W.-B.}\ \bibnamefont {He}}, \bibinfo {author} {\bibfnamefont {S.-J.}\ \bibnamefont {Wu}}, \bibinfo {author} {\bibfnamefont {C.-F.}\ \bibnamefont {Li}}, \ and\ \bibinfo {author} {\bibfnamefont {G.-C.}\ \bibnamefont {Guo}},\ }\href {\doibase 10.1103/PhysRevLett.131.050801} {\bibfield  {journal} {\bibinfo  {journal} {Physical Review Letters}\ }\textbf {\bibinfo {volume} {131}},\ \bibinfo {pages} {050801} (\bibinfo {year} {2023})}\BibitemShut {NoStop}%
\bibitem [{\citenamefont {Wang}\ and\ \citenamefont {Ma}(2025)}]{WangMa2025}%
  \BibitemOpen
  \bibfield  {author} {\bibinfo {author} {\bibfnamefont {X.}~\bibnamefont {Wang}}\ and\ \bibinfo {author} {\bibfnamefont {G.}~\bibnamefont {Ma}},\ }\href {\doibase 10.1088/0256-307X/42/1/014301} {\bibfield  {journal} {\bibinfo  {journal} {Chinese Physics Letters}\ }\textbf {\bibinfo {volume} {42}},\ \bibinfo {pages} {014301} (\bibinfo {year} {2025})}\BibitemShut {NoStop}%
\bibitem [{\citenamefont {Zhou}\ and\ \citenamefont {Zhang}(2005)}]{ZhouHaiJun2005PRL}%
  \BibitemOpen
  \bibfield  {author} {\bibinfo {author} {\bibfnamefont {H.}~\bibnamefont {Zhou}}\ and\ \bibinfo {author} {\bibfnamefont {Y.}~\bibnamefont {Zhang}},\ }\href {\doibase 10.1103/PhysRevLett.94.028104} {\bibfield  {journal} {\bibinfo  {journal} {Phys. Rev. Lett.}\ }\textbf {\bibinfo {volume} {94}},\ \bibinfo {pages} {028104} (\bibinfo {year} {2005})}\BibitemShut {NoStop}%
\bibitem [{\citenamefont {Souslov}\ \emph {et~al.}(2019)\citenamefont {Souslov}, \citenamefont {Dasbiswas}, \citenamefont {Fruchart}, \citenamefont {Vaikuntanathan},\ and\ \citenamefont {Vitelli}}]{PhysRevLett.122.128001}%
  \BibitemOpen
  \bibfield  {author} {\bibinfo {author} {\bibfnamefont {A.}~\bibnamefont {Souslov}}, \bibinfo {author} {\bibfnamefont {K.}~\bibnamefont {Dasbiswas}}, \bibinfo {author} {\bibfnamefont {M.}~\bibnamefont {Fruchart}}, \bibinfo {author} {\bibfnamefont {S.}~\bibnamefont {Vaikuntanathan}}, \ and\ \bibinfo {author} {\bibfnamefont {V.}~\bibnamefont {Vitelli}},\ }\href {\doibase 10.1103/PhysRevLett.122.128001} {\bibfield  {journal} {\bibinfo  {journal} {Phys. Rev. Lett.}\ }\textbf {\bibinfo {volume} {122}},\ \bibinfo {pages} {128001} (\bibinfo {year} {2019})}\BibitemShut {NoStop}%
\end{thebibliography}


\end{document}